\documentclass[journal=jacsat,manuscript=article]{achemso}
\setkeys{acs}{articletitle = true}
\usepackage[version=3]{mhchem} % Formula subscripts using \ce{}

\usepackage{amsmath}
\usepackage{amsfonts}
\usepackage{lscape}  % Landscape
\usepackage{rotating} % Rotate (also landscape)
\usepackage{siunitx}
\usepackage{multirow}
\usepackage{multicol}
\usepackage[utf8]{inputenc}
\usepackage{tablefootnote}
\usepackage{color}
\usepackage[ruled]{algorithm2e}
\usepackage{booktabs}
\usepackage{xfrac}
\usepackage{xcolor}
\usepackage{threeparttable}
\newcommand{\revS}[1]{\textcolor{black}{#1}}
\newcommand{\noteS}[1]{\textcolor{black}{#1}}
\newcommand{\revE}[1]{\textcolor{black}{#1}}
\newcommand{\revL}[1]{\textcolor{black}{#1}}
\newcommand{\revH}[1]{\textcolor{black}{#1}}
\newcommand{\HF}{\mathrm{HF}}

\newcommand{\MO}{\mathrm{MO}}
\newcommand{\Hsim}{\exp(-X)H\exp(X)}
\newcommand{\HTone}{\tilde{H}}
\newcommand{\TF}{\tilde{F}}
\newcommand{\Tg}{\tilde{g}}
\newcommand{\TL}{\tilde{L}}
\newcommand{\x}[2]{x^{#1}_{#2}}
\renewcommand{\t}[2]{t^{#1}_{#2}}
\newcommand{\s}[2]{s^{#1}_{#2}}
\renewcommand{\u}[2]{u^{#1}_{#2}}
\newcommand{\etal}{\textit{et al.}}

\renewcommand{\b}{\boldsymbol}
\newcommand{\mrm}{\mathrm}
\newcommand{\ket}[1]{\vert #1 \rangle}

\newcommand{\bra}[1]{\langle #1 \vert}

\newcommand{\Dbraket}[2]{\langle #1 \hspace{.10em} \vert \hspace{.10em}  #2 \rangle}

\newcommand{\Tbraket}[3]{\langle #1 \hspace{.10em} \vert \hspace{.10em} #2 \hspace{.10em} \vert \hspace{.10em} #3 \rangle}

\author{Sarai Dery Folkestad}
\affiliation{Department of Chemistry, Norwegian University of Science and Technology, N-7491 Trondheim, Norway}
\author{Eirik Fadum Kj{\o}nstad}
\affiliation{Department of Chemistry, Norwegian University of Science and Technology, N-7491 Trondheim, Norway}
\author{Linda Goletto}
\affiliation{Department of Chemistry, Norwegian University of Science and Technology, N-7491 Trondheim, Norway}
\author{Henrik Koch}
\affiliation{Scuola Normale Superiore, Piazza dei Cavaleri 7, 56126 Pisa, Italy}
\alsoaffiliation{Department of Chemistry, Norwegian University of Science and Technology, N-7491 Trondheim, Norway}
\email{henrik.koch@sns.it}  

\title{Multilevel CC2 and CCSD in reduced orbital spaces: electronic excitations in large molecular systems
}

\abbreviations{IR,NMR,UV}

\begin{document}


\begin{abstract}
We present efficient implementations of the multilevel CC2 (MLCC2) and multilevel CCSD (MLCCSD) models. 
As the system size increases, MLCC2 and MLCCSD exhibit the scaling of the lower-level coupled cluster model.
In order to treat large systems, we combine MLCC2 and MLCCSD with a \revE{reduced-space} approach \revE{in which} the multilevel coupled cluster calculation is performed in a significantly truncated molecular orbital basis. 
The truncation scheme is based on the selection of an active region of the molecular system and the subsequent construction of localized Hartree-Fock orbitals. These orbitals are used in the multilevel coupled cluster calculation.
The electron repulsion integrals are  Cholesky decomposed using a screening protocol that guarantees accuracy in the truncated molecular orbital basis \revE{and reduces computational cost}.
The Cholesky factors are constructed directly in the truncated basis, ensuring low storage requirements.
\revE{Systems for which Hartree-Fock is too expensive} can be treated by using a multilevel Hartree-Fock reference.
With the reduced-space approach, we can handle systems with more than a thousand atoms. This is demonstrated for paranitroaniline in aqueous solution.
\end{abstract}

\section{Introduction}
%%%%%%%%%%%%%%%%%%%%%%%%%%%%%%%%
%%%%%%%%%%%%%%%%%%%%%%%%%%%%%%%%
The scaling properties of the coupled cluster hierarchy 
of methods
severely limits the systems for which it is applicable.\citep{helgaker2014molecular} 
The methods have polynomial computational scaling, $\mathcal{O}(N^n)$, 
where $N$ is a measure of system size and $n$ increases with \revE{accuracy} of the method. 
Memory and disk space requirements also increase significantly as one moves up through the hierarchy.

The development of reduced cost and reduced scaling coupled cluster methods has been an active topic for decades.
Arguably, 
the most popular approach has emerged from the work of Pulay and S{\ae}b{\o}.\citep{pulay1983,saebo1993}
They demonstrated that dynamical electronic correlation could be \revE{compactly described} using localized orbitals rather than canonical orbitals; specifically, they used localized occupied \revE{molecular} orbitals \revE{(MOs)}, such as Boys\citep{boys1960construction} or Pipek-Mezey\citep{pipek1989fast} orbitals, and projected atomic orbitals\citep{pulay1983,saebo1993} (PAOs) to span the virtual space.
\revS{Their} local correlation approach was later applied to coupled cluster theory by Hampel, Werner, and Sch{\"u}tz.\citep{hampel1996,schutz2001}
Other local coupled cluster methods include the
local pair natural orbital\citep{neese2009,riplinger2013} and the orbital-specific-virtual\citep{yang2012orbital} coupled cluster methods.
Whereas the success of these local coupled cluster methods in the description of the ground state correlation energy is indisputable,
their extension to excited states has turned out to be more complicated.\citep{korona2003local,kats2006local,kats2009multistate,helmich2013pair,dutta2016towards,dutta2018exploring}

A different approach originates from the multireference coupled cluster method of Oliphant and Adamowicz.\citep{oliphant1991multireference,piecuch1993state,kallay2002general} 
While introduced to describe multireference character, the method is formulated in the framework of single reference coupled cluster theory.
An active orbital space is used, and higher order excitation operators (e.g., triple or quadruple excitations) are included with some indices restricted to the active space.
K\"{o}hn and Olsen\citep{kohn2006coupled} recognized that the method could be used to reduce the cost for single reference systems, and this was further demonstrated by K\'{a}llay and Rolik.\citep{rolik2011cost} 
\revS{The
multilevel coupled cluster (MLCC) approach, introduced by Myhre \etal,\citep{myhre2013extended, myhre2014mlcc, myhre2016multilevel} is closely related to this active space approach}.

In MLCC,
the goal is to accurately describe excitation energies and other intensive properties, rather than extensive properties such as correlation energies. 
This is done by restricting the higher order excitation operators to excite within an active orbital space. 
For example, 
in the multilevel CCSD (MLCCSD)\citep{myhre2013extended,folkestad2019multilevel} method, the double excitation operator is restricted to excite out of active occupied orbitals and into active virtual orbitals.
In this work, \revE{we demonstrate the available computational savings of the multilevel CC2 (MLCC2) and MLCCSD models introduced in Ref.~\citenum{folkestad2019multilevel}; for sufficiently large inactive spaces, we show that the cost is dominated by the lower-level method. 
This has previously been demonstrated for multilevel CC3 (MLCC3) by Myhre \etal\citep{myhre2016multilevel}}

The scaling of the lower-level model \revS{cannot}, however, be avoided. Therefore, in order to use \revL{these} methods for large systems, they must be combined with other multilevel or multiscale approaches. 
For instance, MLCC could be used within a QM/MM\citep{warshel1972,levitt1975} framework 
or with the polarizable continnum model.
\citep{tomasi2005quantum,mennucci2012polarizable} 
Here, we have chosen \revL{to perform} MLCC calculations in a significantly truncated \revE{MO} basis.
The truncation of the MO basis in coupled cluster calculations is used routinely. For example, the frozen core approximation falls into this category\revE{,} \revS{and there are several examples of truncation of natural orbitals, both of the virtual and occupied spaces.\citep{taube2008frozen,landau2010frozen,deprince2013accurate,deprince2013accuracy,kumar2017frozen,mester2017reduced}
The LoFEx\citep{baudin2016lofex,baudin2017lofex} and CorNFLEx\citep{baudin2017correlated} approaches are also notable reduced space coupled cluster approaches that target accuracy in the excited states. In these approaches, a mixed orbital basis consisting of natural transition orbitals (NTOs)\citep{luzanov1976application,martin2003natural,hoyvik2017} and localized orbitals is used. The active space is expanded until the excitation energies have converged to within a predefined threshold.
One drawback of LoFEx and CorNFLEx is that they are state specific methods, i.e., several subsequent calculations with different truncated MO bases must be performed to obtain \revE{a set of} excitation energies. As a consequence, the calculation of transition moments between excited states is complicated by the fact that the states are non-orthogonal and interacting.
An orbital selection procedure similar to that of LoFEx and CorNFLEx has also been used for reduced scaling second-order algebraic diagrammatic construction (ADC(2)) calculations by Mester \etal\citep{mester2019reduced,mester2017reduced}}

\revS{Here,} we use a truncation scheme for the MOs 
\revS{where}
%based on the construction of 
semi-localized Hartree-Fock orbitals (virtual and occupied)
\revE{are constructed and}
%, which can be 
used to calculate localized intensive properties in large molecular systems.
When the region of interest is sufficiently small compared to the full system, the number of MOs in the coupled cluster calculation is much smaller than the number of atomic orbitals (AOs). 
\revS{This reduced space approach has previously been used with standard coupled cluster models,\citep{eTprogram,sanchez2010cholesky} and a very similar approach has been used together with local coupled cluster models.\citep{mata2008correlation}}
\revL{For sufficiently large systems, the cost of Hartree-Fock can become a limiting factor. When this is the case,
we handle it by combining} \revE{the reduced space MLCC approach with a multilevel} Hartree-Fock\citep{saether2017density,hoyvik2019convergence} (MLHF) reference wave function.

The MLCC2 and MLCCSD implementations are based on Cholesky decomposed electron repulsion integrals.\citep{Beebe1977,Koch2003} \revS{We use the two-step Cholesky decomposition algorithm introduced in Ref.~\citenum{Folkestad2019}. In this algorithm, the Cholesky basis and the Cholesky vectors are determined in two separate steps.}
We \revS{have implemented a} direct construction of the Cholesky vectors in the truncated MO basis. This reduces the memory requirement of the vectors from $\mathcal{O}(N_{\mrm{AO}}^3)$ to $\mathcal{O}(N_{\mrm{AO}}n_{\mrm{MO}}^2)$, making it possible to efficiently perform reduced space calculations on systems with several thousands of basis functions. \revH{We use $n$ as a measure of the size of the active space, which does not scale with the system.}
\revS{It should be noted that storage of the Cholesky vectors in the AO basis, albeit temporary, can only be avoided in a decomposition algorithm that determines the Cholesky basis and the Cholesky vectors in separate steps.}
\revL{In the Cholesky decomposition, we also use the }\revS{MO-screening procedure that was introduced in Ref.~\citenum{Folkestad2019}. \revE{This MO-screening} leads to \revE{fewer} Cholesky vectors, further reducing the memory requirement of the Cholesky vectors to $\mathcal{O}(n_{\mrm{MO}}^3)$.}

%%%%%%%%%%%%%%%%%%%%%%%%%%%%%%%%%%%%%%%%%% 
%%%%%%%%%%%%%%%%%%%%%%%%%%%%%%%%%%%%%%%%%%
\section{Theory}
In coupled cluster theory, the wave function is defined as
\begin{align}
    \ket{\mrm{CC}} = \exp(X)\ket{\mrm{HF}},\quad X = \sum_{\mu} x_\mu \tau_\mu
\end{align}
where $\ket{\mrm{HF}}$ is the Hartree-Fock reference, 
$X$ is the cluster operator, 
$x_\mu$ are cluster amplitudes, 
and $\tau_\mu$ are excitation operators. 
The standard models within the coupled cluster hierarchy are 
defined
by restricting $X$ to include the
excitation operators up to a certain order.
In the CC$n$ models, such as CC2\citep{Christiansen1995} and CC3,\citep{Koch1997} the $n$th order excitations are treated perturbatively.

In the following, \revS{the} indices $\alpha, \beta, \gamma, ...$ and $p,q,r,...$ \revE{refer} to \revS{spatial} atomic and molecular orbitals, respectively, and \revS{the} indices $i,j,k,...$ and $a,b,c,...$ \revS{refer} to occupied and virtual orbitals. 
\revS{The total number of occupied and virtual orbitals are denoted by $N_o$ and $N_v$, respectively, and the number of active occupied and active virtual orbitals are denoted by \revS{$n_o^{\mrm{a}}$} and \revS{$n_v^{\mrm{a}}$}}.
%%%%%%%%%%%%%%%%%%%%%%%%%%%%%%%%%%%%%%%%%%
%%%%%%%%%%%%%%%%%%%%%%%%%%%%%%%%%%%%%%%%%%
\subsection{Multilevel CC2 and CCSD}
The MLCC2 cluster operator is given by
\begin{align}
    X^{\mrm{MLCC2}} = X_1 + S_2,
\end{align}
where the single excitation operator, $X_1$, is 
unrestricted, i.e.~defined for all orbitals, whereas the double excitation operator, $S_2$, is restricted to excite within an active orbital space.
As in standard CC2, $S_2$ is treated perturbatively. 
The MLCC2 ground state equations are given by
\begin{align}
        \Omega_{\mu_1} = &\Tbraket{\mu_1}{\hat{H} + [\hat{H}, S_2]}{\mrm{HF}} = 0\label{eq:mlcc2_1}\\
        \Omega_{\mu_2^{S}} = &\Tbraket{\mu_2^{S}}{\hat{H} + [F, S_2]}{\mrm{HF}} = 0\label{eq:mlcc2_2},
\end{align}
where $\hat{H}$ is the $X_1$-transformed Hamiltonian. The doubles projection space, $\{\bra{\mu_2^{S}}\}$, is associated with $S_2$. Except for the restriction of $S_2$ and the projection space, these equations are equivalent to the standard CC2 ground state equations. 
The MLCC2 equations are solved in a basis where the active-active blocks of the occupied-occupied and virtual-virtual Fock matrices are diagonal. In this \textit{semicanonical} basis, eq \eqref{eq:mlcc2_2} can be solved analytically for the $S_2$ amplitudes \revS{in each iteration.
The double amplitudes are inserted into eq \eqref{eq:mlcc2_1}, which is solved with a DIIS-accelerated\citep{pulay1980convergence} quasi-Newton solver\citep{scuseria1986accelerating} to obtain $X_1$.}
\revS{
The MLCC2 equations are formulated in terms of the Cholesky vectors in the $X_1$-basis. See Appendix A for detailed expressions.
}

\revS{
If we consider a fixed active space, the overall scaling of the MLCC2 ground state equations is $\mathcal{O}(N^4)$: the $X_1$-transformation of the Cholesky vectors scales as $\mathcal{O}(N^4)$, as does the computation of the Fock matrix in the $X_1$-basis and the correlation energy. 
The construction of $\b\Omega$ scales as $\mathcal{O}(N^2)$.
}

The MLCC2 excitation energies are determined as the eigenvalues of the Jacobian matrix,
\begin{equation}
  \b A^{\mrm{MLCC2}} = 
  \begin{pmatrix}
    \Tbraket{\mu_1}{[\hat{H},\tau_{\nu_1}]+ [[\hat{H}, \tau_{\nu_1}], S_2]}{\mrm{HF}} &
    \Tbraket{\mu_1}{[\hat{H}, \tau_{\nu_2^{S}}]}{\mrm{HF}}  \\
    \Tbraket{\mu_2^{S}}{[\hat{H}, \tau_{\nu_1}]}{\mrm{HF}} &
    \Tbraket{\mu_2^{S}}{[F, \tau_{\nu_2^{S}}]}{\mrm{HF}} 
  \end{pmatrix}.
\end{equation}
\revS{Here, $\tau_{\nu_2^S}$ is a double excitation included in $S_2$.}
The excited state equations also assume the same form as in standard CC2, except for the restrictions of $S_2$, and the same strategies can therefore be used to solve the MLCC2 equations.\citep{Christiansen1995,hattig2000cc2}
\revS{The most expensive term in the transformation by $\b{A}^\mrm{MLCC2}$ appear at the CCS level of theory (see Appendix A); these terms scale as $\mathcal{O}(N^4)$ and no indices are restricted to the active space. Thus, the overall scaling is $\mathcal{O}(N^4)$.}

In MLCCSD, 
one defines two sets of active orbitals,
where one is a subset of the other. 
The cluster operator has the form
\begin{align}
    X^{\mrm{MLCCSD}} = X_1 + S_2 + T_2, 
\end{align}
where $X_1$ is unrestricted, $S_2$ is restricted to the larger active orbital space, and $T_2$ is restricted to the smaller active orbital space. 
The $S_2$ operator is treated perturbatively (as in CC2 and MLCC2) and
$T_2$ acts as a correction to $S_2$
in the smaller active space. 
This framework is flexible\revL{, since} it allows for
both two-level calculations (CCS/CCSD and CC2/CCSD) and three-level calculations (CCS/CC2/CCSD). 
Previously, we have found that the cheaper and significantly simpler CCS/CCSD method performs very well.\citep{folkestad2019multilevel} 
In the CCS/CCSD method, 
the MLCCSD cluster operator reduces to
\begin{align}
    X^{\mrm{MLCCSD}} = X_1 + T_2,
\end{align}
and only the active space for $T_2$ is needed. In this work, we only consider the CCS/CCSD method.

The MLCCSD (CCS/CCSD) ground state equations are
\begin{align}
    \Omega_{\mu_1} = &\Tbraket{\mu_1}{\hat{H} + [\hat{H}, T_2]}{\mrm{HF}} = 0\label{eq:mlccsd_1}\\
    \Omega_{\mu_2^{T}} = &\Tbraket{\mu_2^{T}}{\hat{H} + [\hat{H}, T_2] + \frac{1}{2} [[\hat{H}, T_2], T_2]}{\mrm{HF}} = 0,\label{eq:mlccsd_2}
\end{align}
where the doubles projection space, $\{\bra{\mu_2^{T}}\}$, is associated with $T_2$.
Equations \eqref{eq:mlccsd_1} and \eqref{eq:mlccsd_2} are equivalent to the standard CCSD equations, except for the restriction of
$T_2$ and the projection space. 
\revS{The construction of \revE{the singles part of} $\b\Omega$, eq \eqref{eq:mlccsd_1}, has the same cost as constructing the MLCC2 $\b\Omega$ ($\mathcal{O}(N^2)$). When the active space is fixed, the construction of \revE{the doubles part of} $\b\Omega$, eq \eqref{eq:mlccsd_2}, scales as $\mathcal{O}(N)$ due to the calculation of the integrals from the Cholesky vectors; all orbital indices are restricted to the active space.}

The excitation energies are obtained as the eigenvalues of the MLCCSD Jacobian, 
\begin{align}
  &\b A^{\mrm{MLCCSD}} =
  &\begin{pmatrix}
    \Tbraket{\mu_1}{[\hat{H},\tau_{\nu_1}]+ [[\hat{H}, \tau_{\nu_1}], T_2]}{\mrm{HF}} &
    \Tbraket{\mu_1}{[\hat{H}, \tau_{\nu_2^{\mrm{T}}}]}{\mrm{HF}} \\
    \Tbraket{\mu_2^{{T}}}{[\hat{H},\tau_{\nu_1}]+ [[\hat{H}, \tau_{\nu_1}], T_2]}{\mrm{HF}} &
    \Tbraket{\mu_2^{{T}}}{[\hat{H},\tau_{\nu_2^{{T}}}]+ [[\hat{H}, \tau_{\nu_2^{{T}}}], T_2]}{\mrm{HF}}
  \end{pmatrix},
\end{align}
\revS{where, $\tau_{\nu_2^T}$ is a double excitation included in $T_2$.}

\revS{
In addition to the terms of the transformation by $\b{A}^{\mrm{MLCC2}}$ that enter the transformation by $\b{A}^{\mrm{MLCCSD}}$, there are terms which scale as $\mathcal{O}(1)$,  $\mathcal{O}(N)$, and $\mathcal{O}(N^2)$. Integral construction for the different terms scales, depending on the number of restricted indices, as $\mathcal{O}(N)$, $\mathcal{O}(N^2)$, or $\mathcal{O}(N^3)$. See Appendix A for detailed expressions.
}

\subsection{Partitioning the orbital space\label{sec:orbitals}}
Selecting the active orbital space for a multilevel coupled cluster calculation is not trivial. Generally, the canonical Hartree-Fock orbitals must be transformed\revE{---through occupied-occupied and virtual-virtual rotations---to an orbital basis that can be intuitively partitioned.}
In order to determine the type of orbitals to use, both the targeted property
and
the system must be considered. 
There are two main approaches to select the active spaces.
If the property of interest is \revE{adequately described at a lower level of theory, then the information from that lower level} can be \revL{exploited} to partition the orbitals. An example is the use of \revE{correlated NTOs (CNTOs)}.\citep{hoyvik2017,folkestad2019multilevel} 
If the property of interest is spatially localized, then localized or semi-localized orbitals can be \revL{applied}. For instance, 
Cholesky orbitals\citep{aquilante2006fast,sanchez2010cholesky} have been used in multilevel coupled cluster calculations by Myhre \etal\citep{myhre2014mlcc,myhre2016multilevel,myhre2016near}

The \revE{CNTOs} are constructed using excitation vectors, $\b{R}$, from a lower-level calculation. The matrices 
\begin{align}
    M_{ij} = \sum_{a} R_{ai}  R_{aj} + \frac{1}{2} \sum_{abk} (1 + \delta_{ai,bk}\delta_{ij})R_{aibk}R_{ajbk}\label{eq:M}\\
    N_{ab} = \sum_{i}  R_{ai}  R_{bi} + \frac{1}{2} \sum_{ijc} (1 + \delta_{ai,cj}\delta_{ab})R_{aicj}R_{bicj}\label{eq:N}
\end{align}
are diagonalized. The matrices that diagonalize  $\b{M}$ and $\b{N}$ are the transformation matrices of the occupied and virtual orbitals, respectively. From eqs \eqref{eq:M} and \eqref{eq:N} it may seem that the lower level method must include double excitation amplitudes in its parametrization. However, CNTOs can be generated from CCS excitation vectors by constructing approximate double excitation vectors:
\begin{align}
R_{aibj}^{\mrm{CCS}} = -\frac{1}{1+\delta_{ai,bj}}\frac{\bar g_{aibj}}{\epsilon_{ij}^{ab} - \omega^{\mrm{CCS}}}.\label{approx_double_1}
\end{align}
Here, $\omega^{\mrm{CCS}}$ is the CCS excitation energy, and $\epsilon_{ij}^{ab} = \epsilon_a + \epsilon_b - \epsilon_i - \epsilon_j$, where the $\epsilon_q$ are orbital energies. The integrals $\bar g_{aibj}$ are defined as
\begin{align}
  \bar g_{aibj} = \mathcal{P}_{ij}^{ab}\Big(\sum_{cJ} R_{ci}L_{bj}^{J}L_{ac}^J - \sum_{kJ} R_{bk}L_{kj}^{J}L_{ai}^{J}\Big), \label{approx_double_2}
\end{align}
where \revS{$g_{pqrs} = \sum_{J} L^{J}_{pq} L^{J}_{rs}$} are the electronic repulsion integrals in the MO basis and $\mathcal{P}_{ij}^{ab} I_{ai,bj} = I_{ai,bj} + I_{bj,ai}$ ($I_{aibj}$ are elements of a rank-4 tensor). 
\revE{Equations} \eqref{approx_double_1} and \eqref{approx_double_2} were suggested by Baudin and Kristensen\citep{baudin2017correlated} \revE{and is based on CIS(D).\citep{head1994doubles}}
\noteS{In our previous work, we have found that the CNTOs obtained from a CCS calculation (using eqs \eqref{approx_double_1} and \eqref{approx_double_2}) perform well, considering accuracy and cost, compared to CNTOs from a CC2 calculation.\citep{folkestad2019multilevel} It should be noted, however, that these orbitals are not expected to perform well for states dominated by double excitations with respect to the reference.}

The active space is selected by considering the eigenvalues of $\b{M}$ and $\b{N}$: \revE{active orbitals result from the eigenvectors corresponding to the largest eigenvalues.}
In this work, we \revS{either explicitly select the number of active occupied and active virtual orbitals ($n_o^{\mrm{a}}$ and $n_v^{\mrm{a}}$) or we} select $n_o^{\mrm{a}}$ and let the number of active virtual orbitals be determined from the total fraction of virtual to occupied orbitals, \revE{i.e.,}
\begin{align}
    n_v^{\mrm{a}} = \frac{N_v}{N_o}n_o^{\mrm{a}}.\label{eq:select_CNTOs}
\end{align}
Alternatively, one can use the selection criterion given in Ref.~\citenum{hoyvik2017}. \revS{This latter approach is more suitable for production calculations; on the other hand, eq \eqref{eq:select_CNTOs} is convenient for testing of the models.}
\revE{Several excited states can be considered simultaneously by diagonalizing sums of $\b M$ and $\b N$ matrices generated from the individual excitation vectors (eqs \eqref{eq:M} and \eqref{eq:N}).\citep{folkestad2019multilevel}}

Cholesky orbitals\citep{aquilante2006fast,sanchez2010cholesky} are obtained by a \revE{restricted} Cholesky decomposition of the Hartree-Fock densities (occupied and virtual); the pivots of the decomposition procedure are restricted to correspond to AOs centered on active atoms.

As an alternative to Cholesky orbitals for the virtual space, one can use projected atomic orbitals\citep{saebo1993} (PAOs).
To construct the PAOs,
the occupied orbitals are projected out of the AOs, $\{\chi_\alpha\}$, centered on the active atoms:
\begin{align}
    \begin{split}
    \chi_\alpha^{\text{PAO}} &= \chi_\alpha - \sum_i\Dbraket{\phi_i}{\chi_\alpha}\phi_i\\
    & = \chi_\alpha - \sum_{i\beta\gamma}C_{\beta i}C_{\gamma i}\Dbraket{\chi_\beta}{\chi_\alpha}\chi_\gamma\\
    &= \chi_\alpha - \sum_{\gamma} \chi_{\gamma}[\b{D}\b{S}]_{\gamma\alpha}.
    \end{split}\label{eq:PAOs}
\end{align}
Here, $\b C$ is the orbital coefficient matrix, $\b{D}$ is the idempotent Hartree-Fock density, and $\b S$ is the AO overlap matrix. 
The orbital coefficient matrix for the active PAOs is therefore $\b C^{\text{PAO}} = \b I - \b{DS'}$, where $\b S'$ is rectangular and contains the columns of $\b S$ which correspond to AOs on active atomic centers. The PAOs are non-orthogonal and linearly dependent. In order to remove linear dependence and orthonormalize the active virtual orbitals, we use the L\"{o}wdin canonical orthonormalization procedure.\citep{lowdin1970nonorthogonality} 
The inactive virtual orbitals are obtained in a similar way: 
\revE{the occupied orbitals, as well as the active virtual orbitals,}
are projected out of the AOs and the resulting orbitals are finally orthonormalized.

After the orbitals \revE{have been} partitioned---regardless of which orbitals are used---we transform to the semicanonical MO basis that is used in MLCC2 and MLCCSD calculations. 
\revE{This transformation involves block-diagonalizing}
the virtual-virtual and occupied-occupied Fock matrices such that the active-active and inactive-inactive blocks become diagonal.

\subsection{Reduced space multilevel coupled cluster}
\revE{Recall that}
MLCC methods 
exhibit the scaling of the lower-level \revS{coupled cluster} model. 
\revE{To overcome this limitation, we apply a reduced space approach}
where only a subregion of the molecule is described at the coupled cluster level. 
The orbitals \revE{in} this subregion 
are divided into active and inactive sets for the MLCC calculation.
The rationale behind this approach is that localized intensive properties 
can be described 
by using accurate and expensive methods only for the region of interest. 
\revE{In particular, it is assumed that 
the effect} of the more distant environment 
is
sufficiently well captured
through contributions to the Fock matrix. 
A few numerical results\citep{sanchez2010cholesky,eTprogram} indicate that excitation energies can be described accurately with this \textit{frozen Hartree-Fock} approach. However, a comprehensive study has not \revS{yet} been published.

To perform reduced space \revS{MLCC} calculations, we must first choose the region of the molecular system to be treated with MLCC. After the Hartree-Fock calculation, localized occupied and virtual orbitals are constructed for the active region. 
\revE{Any localization procedure can be \revL{employed}; however, we use}
Cholesky orbitals for the occupied space and PAOs for the virtual space. 
This set of orbitals 
\revS{enters} the \revS{MLCC} calculation. The remaining occupied orbitals enter the equations through their contributions to the Fock matrix, 
\begin{align}
   \label{eff_fock}
   \begin{split}
   F_{pq} &= h_{pq} + \sum_{i=1}^{N_o}\left(2g_{pqii} - g_{piiq}\right) +
   \sum_{I=1}^{N_o^{\mrm{e}}} \left(2g_{pqII} - g_{pIIq}\right)\\
   &= h_{pq} + \sum_{i=1}^{N_o}\left(2g_{pqii} - g_{piiq}\right) + F_{pq}^{\mrm{e}}.
   \end{split}
\end{align}
\revE{Here,} $N_{o}^{\mrm{e}}$ is the number of frozen occupied orbitals and the index $I$ denotes a frozen occupied orbital. 
The multilevel coupled cluster calculation now has $n_{\mrm{MO}} \ll N_{\mrm{AO}}$, but the procedure is otherwise unchanged: the reduced set of MOs is partitioned into active and inactive sets and the MLCC equations are solved. \revH{We write $n_{\mrm{MO}}$ (lower case $n$) to indicate that the number of MOs does not scale with the system in such calculations.}

A multilevel Hartree-Fock\citep{saether2017density,hoyvik2019convergence} (MLHF) reference  
can also be used.
As in MLCC, 
one first determines the active orbitals:
a set of active atoms 
is
selected, and the active occupied orbitals are obtained through a partial limited Cholesky decomposition of the initial idempotent density; PAOs can be used to determine the active virtual orbitals.
\revE{Only the active orbitals are optimized in the Roothan-Hall procedure, which is}
performed in the MO basis.\citep{hoyvik2019convergence} 
The inactive orbitals enter the optimization through an \textit{effective} Fock matrix that assumes the same form as in eq \ref{eff_fock}.
The inactive two-electron contribution ($\b F^{\mrm{e}}$) is only computed once at the beginning of the calculation and is subsequently transformed to the updated MO basis in every iteration. For details, see Ref.~\citenum{saether2017density}.

\begin{figure*}
    \centering
    \includegraphics[width=0.85\textwidth]{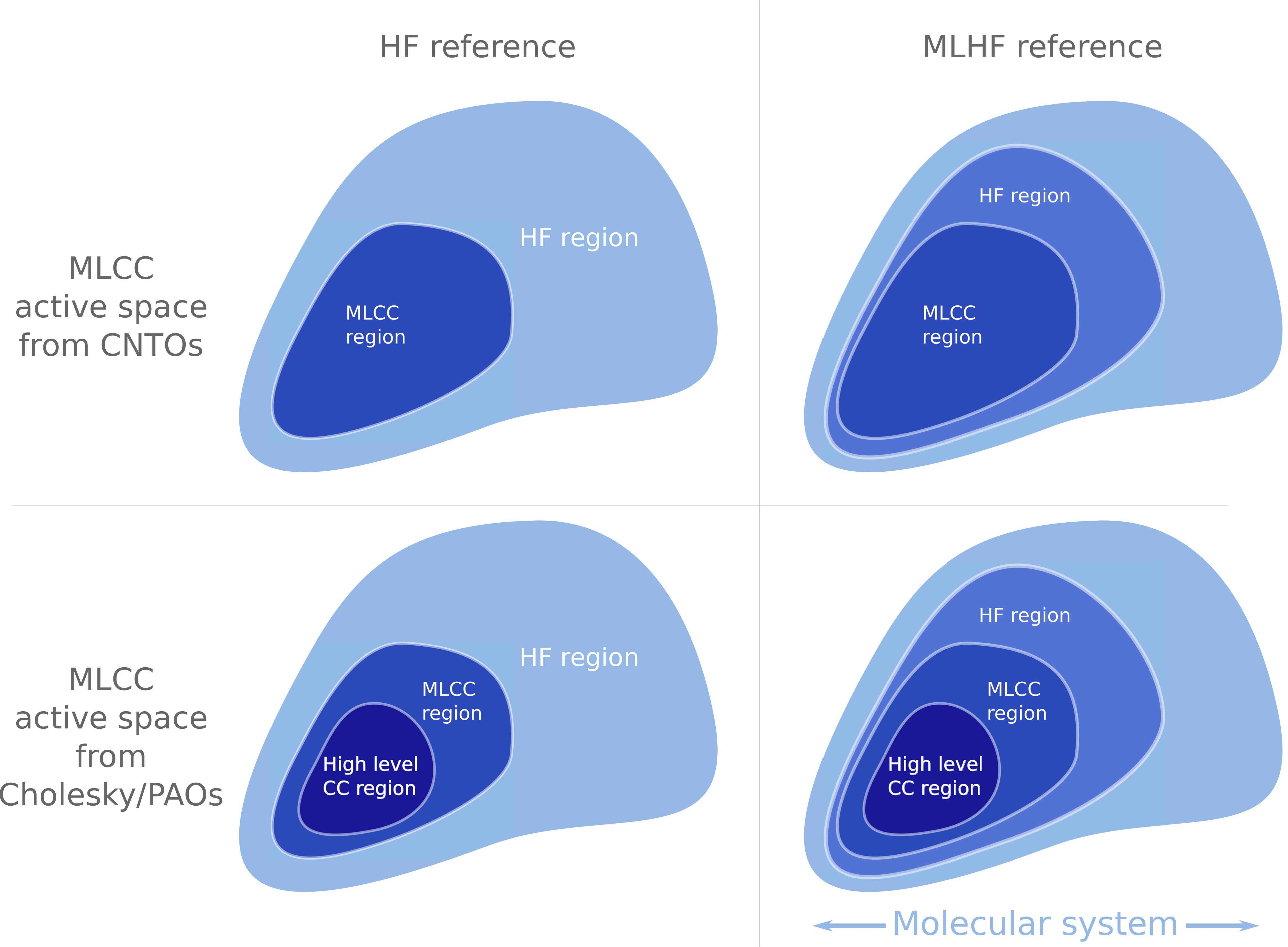}
    \caption{The different levels of active atoms used in
    reduced space MLCC calculations. Left panels show active atoms configurations of reduced space MLCC calculation with an HF reference. Right panels show active atom configurations of reduced space MLCC calculation with an MLHF reference. The two lower panels show the active atom configurations when Cholesky/PAOs are used to determine the active orbitals of the MLCC calculation.}
    \label{fig:active_spaces}
\end{figure*}

The reduced space MLCC approach relies on the definition of levels of active regions of the system, see Figure \ref{fig:active_spaces}.
We must first select which atoms are active in the Hartree-Fock (HF) calculation. If all atoms are active, we have a standard HF reference. 
Secondly, we must determine which atoms enter the MLCC calculation. Lastly, if we use Cholesky/PAOs to partition the orbitals in the MLCC calculation, we must determine which atoms should be treated with the higher level coupled cluster method. This is not necessary when CNTOs are used. Note that the active atom sets for higher level methods are contained within the active atom sets of lower level methods (see Figure \ref{fig:active_spaces}).

Since these methods rely on \revE{selecting} active regions, they are especially well suited for solute/solvent systems. They may
also be used for other large systems where the region of interest is known.

\subsection{Integral handling for reduced space calculations}
\revE{When $n_{\mrm{MO}} \ll N_{\mrm{AO}}$ and $N_{\mrm{AO}}$ is large, as is often the case in reduced space calculations, the electron repulsion integrals must be handled carefully to avoid prohibitive scaling with total system size. In the AO basis,} the Cholesky vectors, $\b{L}^J$, have a storage requirement of $\mathcal{O}(N_{\mrm{AO}}^3)$:
as demonstrated by R{\o}eggen and Wisl{\o}ff-Nilssen,\citep{roeggen1986} the number of Cholesky vectors, $N_J$, is approximately $M N_{\mrm{AO}}$ when a decomposition threshold of $10^{-M}$ is used. 
For example, with a loose decomposition threshold of $10^{-2}$, \revS{about} $28$ TB \revL{of memory} is needed to store the Cholesky vectors of a molecular system with $12000$ AOs\revE{---assuming double precision and no screening.}

We have previously suggested a two-step Cholesky decomposition algorithm\citep{Folkestad2019} in which the Cholesky basis \revE{(i.e., the set of pivots)}, $\mathcal{B}$, is determined in the first step. The Cholesky vectors are constructed in the second step through an RI-like expression,
\begin{align}
    L_{\alpha\beta}^{J} = \sum_{K}(\alpha\beta|K)[\b Q^{-T}]_{KJ},
\end{align}
where the matrix $\b{Q}$ is the Cholesky factor of the matrix $S_{KL} = (K|L)$ for $K,L\in\mathcal{B}$. 
This two-step algorithm 
makes it possible to directly construct the
Cholesky vectors in the MO basis:
\begin{align}
	\begin{split}
    L_{pq}^{J} 
    & = \sum_{\alpha\beta}C_{\alpha p}L_{\alpha\beta}^{J}C_{\beta q}\\
    & = \sum_{\alpha\beta K} C_{\alpha p}C_{\beta q}(\alpha\beta|K)[\b Q^{-T}]_{KJ}.
	\end{split}
\end{align}
We emphasize that \revS{it is not possible \revS{to avoid storing the AO Cholesky vectors} with a \revS{\textit{one-step}} Cholesky decomposition of the AO electron repulsion integral matrix.}
\revE{Alternatively, the MO electron repulsion integrals can be constructed from the AO integrals. To reduce the scaling, one can combine screening on the AO integrals and the MO-coefficients.}

Below we outline an algorithm to construct and store the vectors directly in the MO basis (see Algorithm \ref{alg:cholesky_vectors}). This is done after the elements of the basis $K\in\mathcal{B}$ have been determined, $\b{S}$ has been constructed and decomposed, and $\b{Q}$ has been inverted. 
When the \revE{MO} Cholesky factor, $\b{L}$, is too large to store in memory, $L_{pq}^J$ is constructed for a maximum number of $p$ indices (resulting in several batches, $P_1, P_2, \ldots, P_n$).
The direct construction of the Cholesky vectors in the MO basis reduces the storage requirement to $\mathcal{O}(N_{\mrm{AO}}n_{\mrm{MO}}^2)$. Note that this is linear, rather than cubic, in $N_{\mrm{AO}}$.

\revS{Algorithm \ref{alg:cholesky_vectors} is designed to avoid the IO operations involved in temporary storage and reordering of the intermediate $\b{X}$. Alternatively, $\b{X}$ can be constructed and stored on disk before $\b{L}$ is constructed in batches over $p$ or $q$. With the latter approach, the integrals $(\alpha\beta|K)$ are never recalculated. It should be noted, however, that when $n_\mrm{MO}\ll N_\mrm{AO}$, batching over $p$ is typically not necessary.} 

\vspace{0.4cm}
\begin{algorithm}[H]
\SetAlgoLined
\DontPrintSemicolon
    \KwIn{$\mathcal{B}$, $\b{Q}^{-T}$}
    Determine batches of $p$, $P_1, P_2, \ldots, P_n$\;
    \For {$P_i$}{
        Allocate $X_{pqK}$, $\forall\;p\in P_i$\;
        \For{$K \in \mathcal{B}$}{
            Calculate $(\alpha\beta| K)$\;
            $C_{\alpha p} C_{\alpha q}(\alpha\beta| K)\rightarrow X_{pqK}$, $\forall\;p\in P_i$\;
        }
        $X_{pqK} [\b{Q}^{-T}]_{KJ}\rightarrow L^{J}_{pq}$, $\forall\;p\in P_i$\;
        Store $L^{J}_{pq}$, $\forall\;p\in P_i$\;
    }
 \caption{Constructing MO Cholesky vectors from the RI expression}
 \label{alg:cholesky_vectors}
\end{algorithm}
\vspace{0.4cm}

\revE{The number of Cholesky vectors, $N_J$, can---through a method-specific screening---be made to scale with $n_{\mrm{MO}}$ rather than $N_{\mrm{AO}}$. Consequently, the storage requirements become $\mathcal{O}(n_{\mrm{MO}}^3)$.}
\revE{Method-specific decompositions were first considered by Boman \etal\citep{Boman2008}}
\revS{We use} the active space screening given in Ref.~\citenum{Folkestad2019}. 
In a given iteration of the Cholesky decomposition procedure, 
the next element of the basis is determined by considering the updated diagonal of the integral matrix
\begin{align}
    D_{\alpha\beta} = g_{\alpha\beta\alpha\beta} - \sum_{J\in\mathcal{B}} ({L_{\alpha\beta}^J})^2.
\end{align}
Here, the sum is over the current elements of the basis.
In the standard decomposition algorithm, the next element of the basis is selected as the $K=\alpha\beta$ corresponding to the largest element of $\b{D}$. The decomposition procedure 
is terminated when
\begin{align}
    \max_{\alpha\beta} D_{\alpha\beta} < \tau,\label{eq:standard_cholesky}
\end{align}
where $\tau$ is the decomposition threshold.
\revS{
In the spirit of method specific Cholesky decomposition,\citep{Boman2008} one can consider the Cholesky decomposition of the matrix with elements
\begin{align}
    M_{\alpha\beta pq,\gamma\delta rs} = C_{\alpha p}^{\mrm{a}} C_{\beta q}^{\mrm{a}} g_{\alpha\beta\gamma\delta} C_{\gamma r}^{\mrm{a}} C_{\delta s}^{\mrm{a}}.
\end{align}
The positive semi-definiteness of $\b{M}$ follows directly from the positive semi-definiteness of $\b{g}$.
The diagonal of $\b{M}$,
\begin{align}
    M_{\alpha\beta pq,\alpha\beta pq} = C_{\alpha p}^{\mrm{a}} C_{\beta q}^{\mrm{a}} D_{\alpha\beta}C_{\alpha p}^{\mrm{a}} C_{\beta q}^{\mrm{a}},
\end{align}
is bound from above by
\begin{align}
    M_{\alpha\beta pq,\alpha\beta pq} \leq v_{\alpha}v_{\beta} D_{\alpha\beta},
\end{align}
} where
\begin{align}
    \nu_{\alpha} = \max_{p} {(C_{\alpha p}^{\mrm{a}}})^2,\label{eq:mo_screening}
\end{align}
\revS{and where $\b C^{\mrm{a}}$ is the MO coefficient matrix of the reduced space MLCC calculation. W}e can modify the procedure to determine the Cholesky basis. 
The selection and termination criteria are changed by considering
the screened diagonal,
\begin{align}
   \tilde{D}_{\alpha\beta} = \nu_{\alpha} D_{\alpha\beta}\nu_{\beta},\label{eq:cd_screening}
\end{align}
instead of $\b{D}$.
Using eq \eqref{eq:cd_screening}, we obtain a smaller Cholesky basis compared to the standard decomposition. 
\revS{The MO integrals are, thus, given by
\begin{align}
    g_{pqrs} = \sum_{\alpha\beta\gamma\delta}  (\sum_J L_{\alpha\beta pq}^J L_{\gamma\delta rs}^J + \Delta_{\alpha\beta pq,\gamma\delta rs}),
\end{align}} 
where the errors $\Delta_{\alpha\beta pq,\gamma\delta rs}$ are less than $\tau$.

\revE{Finally, let us briefly consider the computational scaling of the decomposition procedure. Except for the initial integral cutoff screening, which scales as $\mathcal{O}(N_{\mrm{AO}}^2)$ in our implementation, the MO-screened decomposition algorithm scales as $\mathcal{O}(n_{\mrm{MO}}^3)$. 
The prescreening step can be implemented with a lower scaling; however, this step is not time-limiting in any of the reported calculations.}

%%%%%%%%%%%%%%%%%%%%%%%%%%%%%%%%%%%%%%%%%%%%%%
%%%%%%%%%%%%%%%%%%%%%%%%%%%%%%%%%%%%%%%%%%%%%%

\section{Results and discussion}
The MLCC2 and MLCCSD \revE{methods} have been implemented in a development version of the e$^T$ program.\citep{eTprogram}
\revS{The following thresholds are \revL{applied,} unless otherwise stated:
the Hartree-Fock equations are solved to within a gradient threshold of $10^{-8}$; the Cholesky decomposition threshold is $10^{-3}$; the coupled cluster amplitude equations are solved such that $|\b{\Omega}| < 10^{-6}$; the excited state equations are solved to within a residual threshold of $10^{-4}$;
and} occupied Cholesky orbitals are constructed using a threshold of $10^{-2}$ \revS{on the pivots}.
The frozen core approximation is used \revS{throughout.}
All geometries are available from Ref.~\citenum{geometries}.

\subsection{Performance and scaling}
\begin{figure}
 \centering
 \includegraphics[width=0.8\linewidth]{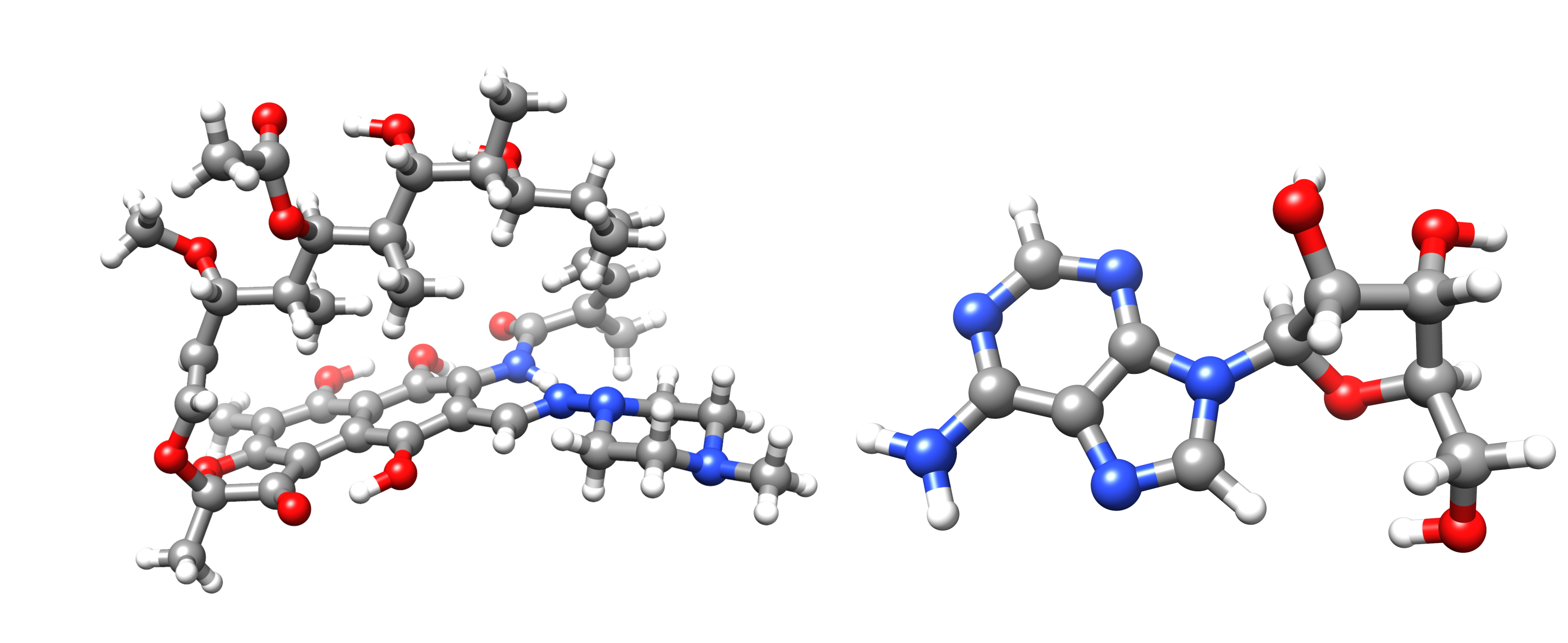} 
 \caption{Rifampicin on the left and adenosine on the right.}
 \label{fig:rifampicin_adenosine}
\end{figure}
\begin{table}[]
    \centering
    \begin{tabular}{l c c c c c c c}
    \toprule
    Method 
    & $n_o^{\text{a}}$ 
    & $n_v^{\text{a}}$ 
    & $\omega~[\si{\eV}]$ 
    & $t^{\text{gs}}~[\si{\hour}]$ 
    & $t^{\text{es}}~[\si{\hour}]$ 
    & $t^{\mrm{CNTO}}~[\si{\hour}]$ 
    & PMU [GB]\\
    \midrule
    \multirow{3}{*}{MLCC2}
    & 40 & 400 & 2.78 & 0.3 & 0.9 & 1.9 & 500.0 \\
    & 60 & 600 & 2.65 & 0.5 & 4.8 & 2.0 & 500.0 \\
    & 80 & 800 & 2.59 & 0.9 & 12.3 & 1.9 & 500.0\\
    \midrule
    CC2 & 161 & 1645 & 2.57 & 32.2 & 183.8 & -- & 498.3 \\
    \bottomrule
    \end{tabular}
    \caption{MLCC2/aug-cc-pVDZ and CC2/aug-cc-pVDZ calculations for rifampicin. 
    $n_o^{\text{a}}$ and $n_v^{\text{a}}$ are the number active occupied and virtual orbitals and $\omega$ is the lowest excitation energy. 
    The wall times to solve the ground and excited state equations ($t^{\text{gs}}$ and $t^{\text{es}}$) and to construct the CNTOs ($t^{\mrm{CNTO}}$) are also given.  
    The calculations were performed on two Intel Xeon E5-2699 v4 processors using 44 threads. \revS{The calculations were performed with 500 GB memory available. Peak memory usage (PMU) is given in GB}}
    \label{tab:rifampicin}
\end{table}

\begin{table}[]
    \centering
    \begin{tabular}{l c c c c c c c c c}
    \toprule
    Method &
    $n_o^{\text{a}}$ & 
    $n_v^{\text{a}}$ & 
    $\omega_1~[\si{\eV}]$ &
    $\omega_2~[\si{\eV}]$ &
    $\omega_3~[\si{\eV}]$ &
    $t^{\text{gs}}~[\si{\minute}]$ 
    & $t^{\text{es}}~[\si{\hour}]$ 
    & $t^{\mrm{CNTO}}~[\si{\minute}]$ &
    PMU [GB]\\
    \midrule
    \multirow{3}{*}{MLCCSD}
    & 25 & 225 & 5.26 & 5.37 & 5.41 & 2.3  & 0.8  & 1.9 & 40.3  \\
    & 30 & 270 & 5.25 & 5.36 & 5.41 & 4.9  & 1.8  & 1.9 & 77.2  \\
    & 35 & 315 & 5.25 & 5.35 & 5.41 & 9.9 & 4.0  & 1.8 & 138.0 \\
    \midrule
    CCSD & 51 & 484 & 5.25 & 5.35 & 5.41 & 75.3 & 38.3 & --- &  288.1\\
    \bottomrule
    \end{tabular}
    \caption{MLCCSD/aug-cc-pVDZ and CCSD/aug-cc-pVDZ calculations on adenosine. 
    $n_o^{\text{a}}$ and $n_v^{\text{a}}$ are the number active occupied and virtual orbitals and $\omega_i$ are excitation energies. 
    The wall times to solve the ground and excited state equations ($t^{\text{gs}}$ and $t^{\text{es}}$) and to construct the CNTOs ($t^{\mrm{CNTO}}$) are also given.  
    The calculations were performed on two Intel Xeon Gold 6138 processors with 40 threads and \revS{${355}$ GB memory available. Peak memory usage (PMU) is given in GB}}
    \label{tab:adenosine}
\end{table}
\begin{table}
    \centering
    \begin{tabular}{l c c c c c c c}
    \toprule
     $n_o^{\text{a}}$ & $n_v^{\text{a}}$ & $\omega~[\si{\eV}]$ & $t^{\text{gs}}~[\si{\hour}]$ & $t^{\text{es}}~[\si{\hour}]$ & $t^{\mrm{CNTO}}~[\si{\hour}]$ & PMU [GB]\\
    \midrule
     40 & 400 & 3.04 & 8.5  & 5.6  & 7.5 & 354.5 \\
     50 & 500 & 3.02 & 13.1  & 9.3 & 7.6 & 354.5\\
     60 & 600 & 3.00 & 14.1 & 21.1 & 5.6 & 354.5 \\
    \bottomrule
    \end{tabular}
    \caption{MLCCSD/aug-cc-pVDZ calculations for rifampicin. 
    $n_o^{\text{a}}$ and $n_v^{\text{a}}$ are the number active occupied and virtual orbitals and $\omega$ is the lowest excitation energy. 
    The wall times to solve the ground and excited state equations ($t^{\text{gs}}$ and $t^{\text{es}}$) and to construct the CNTOs ($t^{\mrm{CNTO}}$) are also given.  
    The calculations were performed on two Intel Xeon Gold 6138 processors with 40 threads and \revS{${355}$ GB memory available. Peak memory usage (PMU) is given in GB}}
    \label{tab:rifampicin_mlccsd}
\end{table}

The MLCC2 and MLCCSD methods can be used to obtain excitation energies \revE{of} CC2 and CCSD quality, at significantly reduced cost. 
This is \revE{demonstrated} for rifampicin and adenosine, see Figure \ref{fig:rifampicin_adenosine}.
\revE{For rifampicin, the} lowest excitation energy is calculated at the MLCC2/aug-cc-pVDZ and CC2/aug-cc-pVDZ levels of theory. 
\revE{For adenosine,} 
the three lowest excitation energies are calculated at the MLCCSD/aug-cc-pVDZ and CCSD/aug-cc-pVDZ levels of theory. 
\revE{We have used} CNTOs to partition the orbitals.
The results are given in Tables \ref{tab:rifampicin} and \ref{tab:adenosine}, respectively.
\revE{These show that
t}he error in the MLCCSD and MLCC2 excitation energies with respect to CC2 and CCSD is smaller than the expected error of CC2 and CCSD.\citep{kannar2014benchmarking,kannar2016accuracy} \revE{Furthermore, the cost is drastically reduced in all cases.}

\revS{In Tables \ref{tab:rifampicin}--\ref{tab:rifampicin_mlccsd}, we have given the available memory and peak memory used in these calculations. Note that the calculations may be performed with less memory since the models are implemented with batching for the memory intensive terms.}

The lowest excitation energy of rifampicin was also calculated with MLCCSD/aug-cc-pVDZ, see Table \ref{tab:rifampicin_mlccsd}. Since the system has $1806$ MOs,
a full CCSD calculation would be \revE{demanding; therefore,} we do not present a reference CCSD calculation. 
However, the variation of the excitation energy is less than $\SI{0.05}{\eV}$ for the different active spaces and can therefore be considered converged. 
In our experience, MLCCSD excitation energies converge smoothly to the CCSD values.\citep{folkestad2019multilevel}
Note that the MLCC2 and MLCCSD timings cannot be compared as the calculations were performed on different processors. 

\begin{figure}
 \centering
 \includegraphics[width=\linewidth]{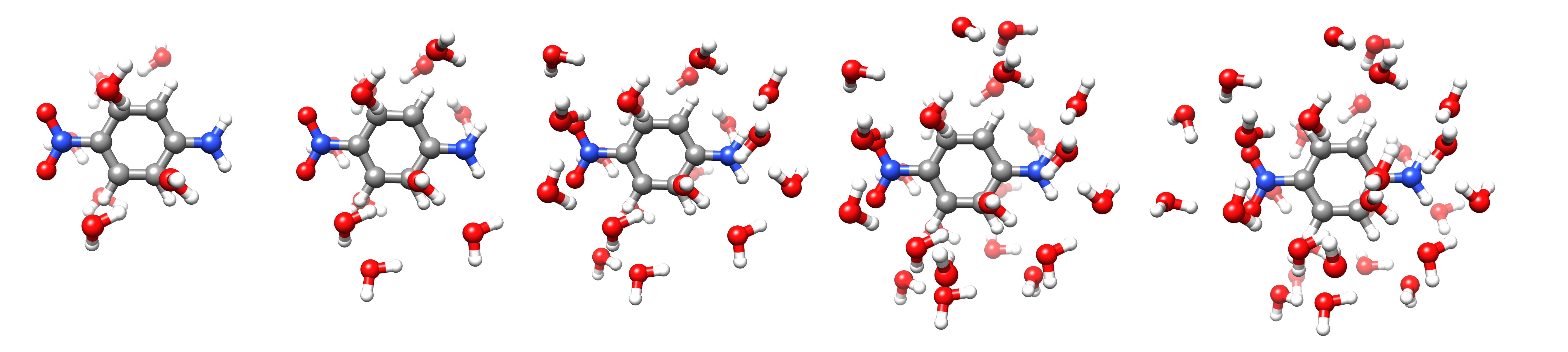} 
 \caption{PNA and water}
 \label{fig:PNA_water_expand}
\end{figure}

\revS{To demonstrate the scaling properties, we consider a system of PNA and water molecules. The size of the active space is fixed\revE{---}with 36 occupied and 247 virtual orbitals\revE{---}and the system size is increased by adding water molecules \revE{(see Figure \ref{fig:PNA_water_expand})}. \revS{We use the} aug-cc-pVDZ basis set. 
}
\begin{figure}
    \centering
    \includegraphics[width=0.95\textwidth]{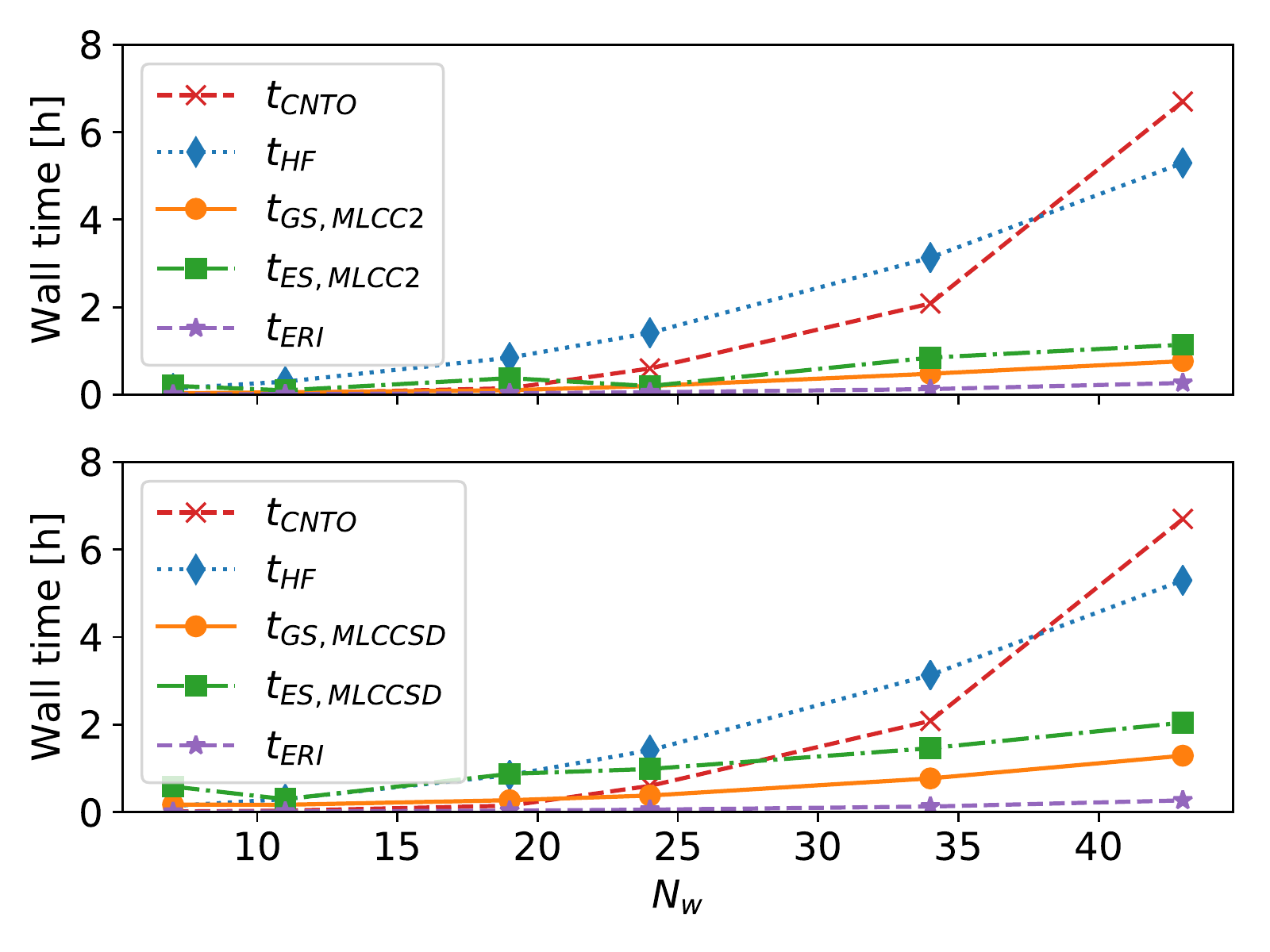}
    \caption{\revS{Timings for MLCC2 (upper) and MLCCSD (lower) calculations on PNA and water. $N_\mrm{w}$ is the number of water molecules, $t_{\mrm{HF}}$ is the full Hartree-Fock calculation time, $t_{\mrm{GS,MLCC2}}$ and  $t_{\mrm{GS,MLCCSD}}$ are the MLCC ground state calculation times,  $t_{\mrm{ES,MLCC2}}$ and $t_{\mrm{ES,MLCCSD}}$ are the MLCC excited state calculation times to obtain a single excited state, $t_{\mrm{ERI}}$ is the time to Cholesky decompose the electron repulsion integrals, and $t_{\mrm{CNTO}}$ is the time to construct the CNTOs. The calculations were performed on two Intel Xeon E5-2699 v4 processors using 44 threads and with 1.4 TB memory available.}}
    \label{fig:scaling_plot1}
\end{figure}

\revS{In Figure \ref{fig:scaling_plot1}, we show the overall wall times of the Hartree-Fock calculation, the CNTO construction, and the MLCC ground and excited state calculations. The steep $\mathcal{O}(N^5)$ scaling of the CNTO construction is apparent: for the largest system, 
it is the most expensive step. The ground and excited state MLCC equations scale as $\mathcal{O}(N^4)$; however, for the larger systems we have considered, the Hartree-Fock calculation is seen to be more expensive. 
This must be understood in the context of system size and the use of an augmented basis set.
For sufficiently large inactive spaces, the $\mathcal{O}(N^4)$ terms of MLCC2 and MLCCSD will become more expensive than Hartree-Fock.
}
\begin{figure}
    \centering
    \includegraphics[width=0.95\textwidth]{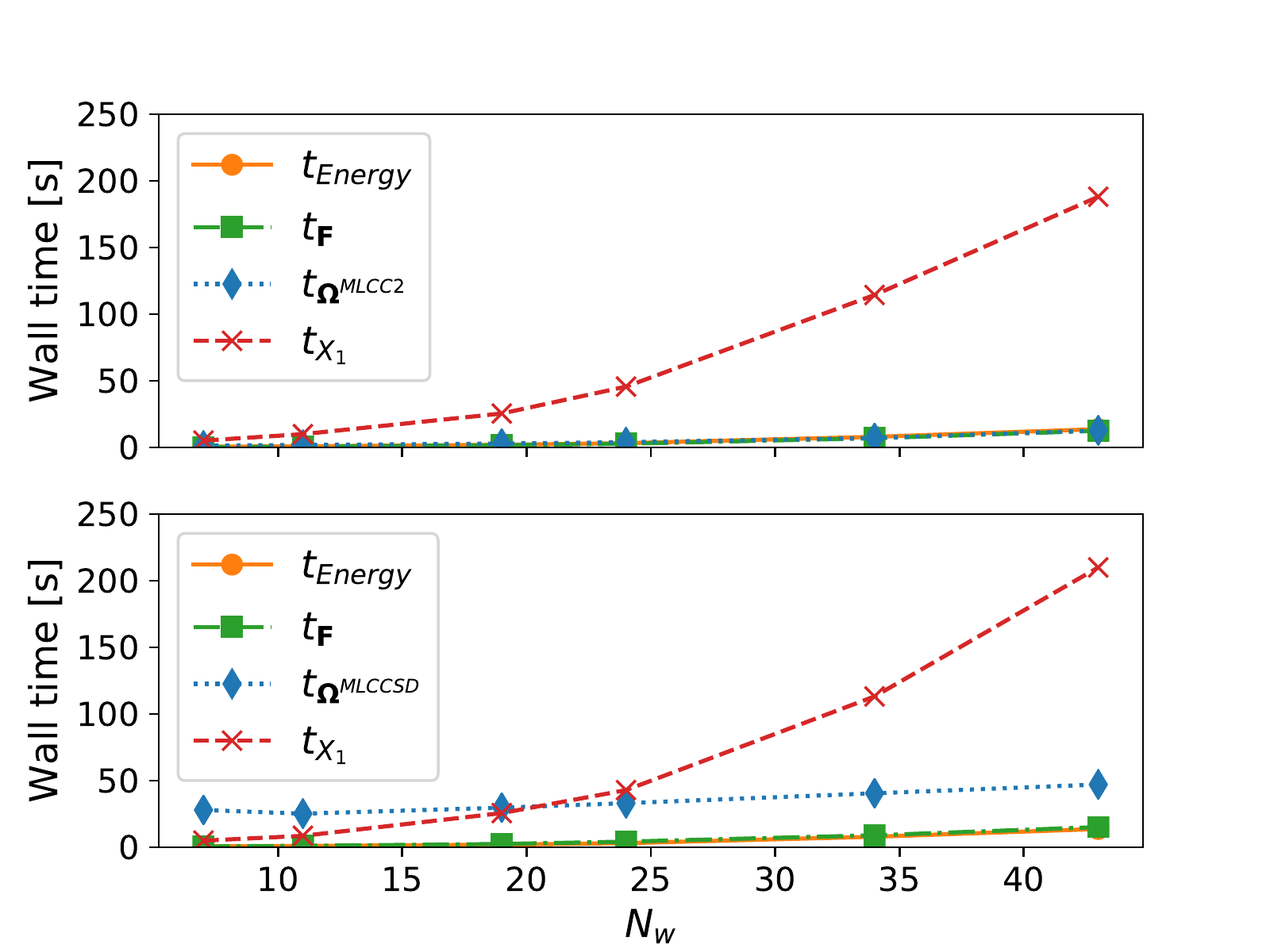}
    \caption{\revS{Timing breakdown of the MLCC2 (upper) and MLCCSD (lower) ground state iteration for PNA and water. $N_\mrm{w}$ is the number of water molecules, $t_{\mrm{Energy}}$ is the time to compute the MLCC correlation energy,  $t_{F}$ is the time to construct the necessary blocks of the Fock matrix in the $X_1$-basis, $t_{\mrm{\Omega}}$ is the time to construct the $\b{\Omega}$-vector, and  $t_{X_1}$ is the time to $X_1$-transform the Cholesky vectors. The calculations were performed on two Intel Xeon E5-2699 v4 processors using 44 threads and with 1.4 TB memory available.}}
    \label{fig:scaling_plot2}
\end{figure}    

\revS{In Figure \ref{fig:scaling_plot2}, we present a timing breakdown of an iteration to solve the MLCC ground state equations. The iteration is dominated by the $\mathcal{O}(N^4)$ step to construct the $X_1$-transformed Cholesky vectors. The calculation of the energy, and the necessary blocks of the Fock matrix in the $X_1$-basis, also scale as $\mathcal{O}(N^4)$, but the prefactor is lower for these operations. The construction of the $\b{\Omega}$-vector scales as $\mathcal{O}(N^2)$. In MLCCSD, the $\b{\Omega}$-vector contains additional contractions, compared to MLCC2, that scale as $\mathcal{O}(1)$ or $\mathcal{O}(N)$ (see Appendix A).}

\begin{figure}
    \centering
    \includegraphics[width=0.95\textwidth]{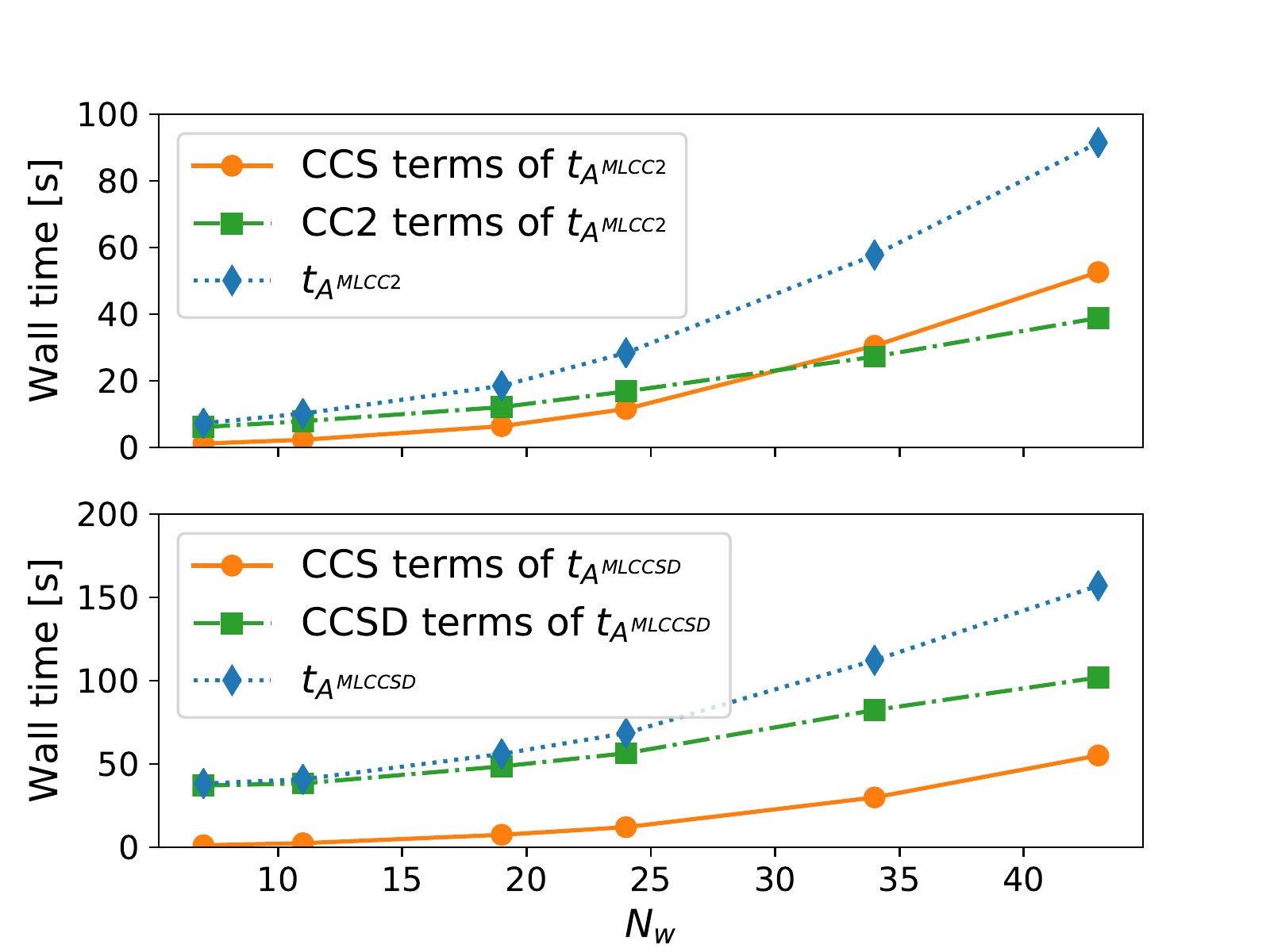}
    \caption{\revS{Wall times of the linear transformation by the MLCC2 (upper) and MLCCSD (lower) Jacobian matrices ($t_{A^{\mrm{MLCC2}}}$ and $t_{A^{\mrm{MLCCSD}}}$) for systems of PNA and water.  The contribution from terms that arise at the CCS level and at the CC2 or CCSD level of theory are plotted separately. $N_\mrm{w}$ is the number of water molecules. The calculations were performed on two Intel Xeon E5-2699 v4 processors using 44 threads and with 1.4 TB memory available.}}
    \label{fig:scaling_plot3}
\end{figure}

\revS{
In Figure \ref{fig:scaling_plot3}, we plot the wall time of the Jacobian matrix transformation together with the time spent on terms that arise at the CCS, CC2, and CCSD level of theory. The CCS-terms scale more steeply ($\mathcal{O}(N^4)$), and for MLCC2, we see that these terms dominate when the inactive space is sufficiently large. For MLCCSD, the CCS-terms are significant, but they do not dominate for any of the systems.
}

\subsection{Reduced space calculations}
\begin{figure}
    \centering
    \includegraphics[width=0.5\linewidth]{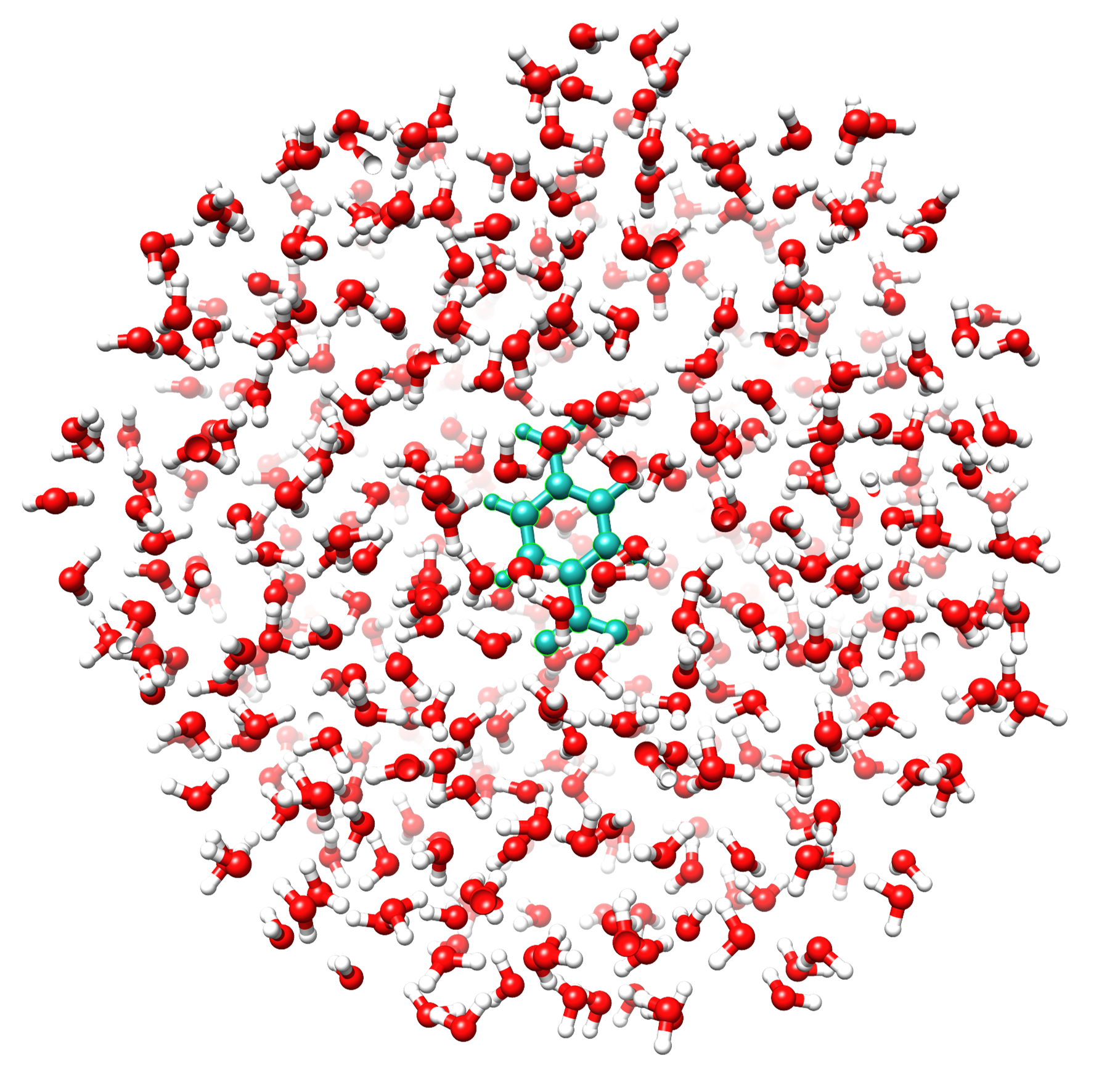}
    \caption{PNA with 499 water molecules.}
    \label{fig:pna_499}
\end{figure}

We now consider a larger PNA-in-water system. The geometry is extracted from a single snapshot of a molecular dynamics simulation taken from Ref.~\citenum{giovannini2019fqfmulinear}.
The PNA-in-water system is restricted to a sphere centered on PNA with a $\SI{15}{\angstrom}$ radius and includes 499 water molecules, see Figure \ref{fig:pna_499}.

\begin{table}
    \centering
    \begin{tabular}{ c c c c c c c c c}
    \toprule
    & \multicolumn{4}{c}{Standard} & \multicolumn{4}{c}{MO-screened} \\\cmidrule(r){2-5}\cmidrule(r){6-9}
    $\tau$ & $N_J$ & $\frac{N_J}{N_{\mrm{AO}}}$ & $\epsilon\;[a.u]$ & $\omega\;[\si{\eV}]$ & $N_J$ &$\frac{N_J}{n_{\mrm{MO}}}$ & $\epsilon\;[a.u]$ & $\omega\;[\si{\eV}]$\\
    \midrule
    $10^{-2}$ & 8434  & 2.1 & $1.1 \cdot 10^{-2}$ & 4.0501 & 606  & 0.9 & 4.77 & No convergence \\
    $10^{-3}$ & 12297 & 3.1 & $1.1 \cdot 10^{-3}$ & 4.0771 & 1440 & 2.2 & 4.77 & 4.1055 \\
    $10^{-4}$ & 15474 & 3.9 & $1.7 \cdot 10^{-4}$ & 4.0761 & 2445 & 3.7 & 4.77 & 4.0785 \\
    $10^{-6}$ & 24826 & 6.3 & $1.6 \cdot 10^{-6}$ & 4.0753 & 5378 & 8.2 & 4.76 & 4.0754 \\
    \bottomrule
    \end{tabular}
    \caption{The lowest MLCCSD-in-HF excitation energy of the PNA-in-water system obtained with regular and MO-screened Cholesky decomposition. The PCD algorithm is used. The threshold, $\tau$, the number of Cholesky vectors, $N_J$, and the largest error in the approximated matrix in the AO basis, $\epsilon$, are given. There are 3971 basis functions.}
    \label{tab:CD_pna_water}
\end{table}
%%%
% DECOMPOSITION TIME: 10^-2 : 515.75s, 10^-3: 865.94s, 10^-4: 1160.69s 10^-6: 2249.69
%
% MO screened: 10^-3: 83.18, 10^-4: 285.85, 10^-6: 289.72 

To \revS{assess the accuracy of the} MO-screening procedure of eqs \eqref{eq:mo_screening} and \eqref{eq:cd_screening}, we consider the lowest MLCCSD-in-HF excitation energy of the system\revL{, which corresponds to} a charge transfer process in PNA. We compare the MO-screened Cholesky decomposition \revE{with} the standard Cholesky decomposition. Note that we use the partitioned Cholesky decomposition (PCD) algorithm, described in Ref.~\citenum{Folkestad2019}, \revE{with two batches}. 
In these MLCCSD calculations, the atoms within a sphere of $\SI{5}{\angstrom}$ are included in the MLCC region ($r_{\mrm{CCS}} = \SI{5}{\angstrom}$) and the atoms within a sphere of radius $\SI{3.5}{\angstrom}$ are defined as active at the CCSD level of theory ($r_{\mrm{CCSD}} = \SI{3.5}{\angstrom}$). The orbitals are partitioned with the Cholesky/PAO approach. 
\revE{For the CCSD/CCS/HF levels of theory, we use the aug-cc-pVDZ/cc-pVDZ/STO-3G basis sets.}  
The total number of basis functions is 3971, and in the MLCCSD-in-HF calculations, we have $n_o^{\mrm{CCSD}} = 90$, $n_v^{\mrm{CCSD}} = 287$, $n_o^{\mrm{CCS}} = 57$, and $n_v^{\mrm{CCS}} = 219$, that is, $n_{\mrm{MO}} = 653$. The results are given in Table \ref{tab:CD_pna_water}. 

The MO-screening yields significantly fewer Cholesky vectors without introducing large errors in the excitation energies. \revE{As expected, the number of Cholesky vectors, $N_J$, is seen to be on the same order of magnitude as $N_{\mrm{AO}}$ and $n_{\mrm{MO}}$ for the standard and MO-screened decomposition algorithms, respectively.}
\revH{Fewer Cholesky vectors reduces the cost of the coupled cluster calculation, where the Cholesky vectors are either used to construct the integrals or applied directly in Cholesky vector-based algorithms. Moreover, the decomposition time is reduced when the MO-screening is employed; for instance, with a threshold of $10^{-4}$, the decomposition time was $\SI{1161}{\second}$ without screening and $\SI{285.85}{\second}$ with screening. In any case, the decomposition time is not a bottleneck in any of these calculations.}

The largest error in the approximated AO integral matrix, $\epsilon$, is also given in Table \ref{tab:CD_pna_water}. 
For standard PCD, 
the
errors are comparable to the decomposition threshold. 
With MO-screening, $\epsilon$ is large because AO integrals that do not contribute to the MO integrals are not described by the Cholesky vectors.
Without MO-screening, a Cholesky decomposition threshold of $10^{-2}$ or $10^{-3}$ is typically sufficient.\citep{Folkestad2019} 
For MLCC2 or MLCCSD in a reduced space calculation, the MO-screening can be used and a threshold of $10^{-4}$ seems suitable. In the calculation with MO-screening and a threshold of $10^{-2}$, the MLCCSD calculation did not converge. 

\begin{table}
    \centering
    \begin{tabular}{l c c c c c c c}
    \toprule
    & &\multicolumn{3}{c}{MLCC2} & \multicolumn{3}{c}{MLCCSD} \\\cmidrule(r){3-5}\cmidrule(r){6-8}
    Reference & 
    $t^{\mrm{Ref}}~[\si{\hour}]$ & 
    $\omega~[\si{eV}]$ & 
    $t^{\mrm{MLCC}}~[\si{\hour}]$ 
    & PMU [GB]
    & $\omega~[\si{eV}]$ 
    & $t^{\mrm{MLCC}}~[\si{\hour}]$ 
    & PMU [GB]\\  
    \midrule
    HF    & \revS{48.1} & $3.821$ & \revS{$3.1$} & \revS{$370$} & $4.075$ & \revS{6.7} & \revS{382}\\
    MLHF  & \revS{33.6} & $3.832$ & \revS{$3.1$} & \revS{$370$} & $4.083$ & \revS{6.8} & \revS{382}\\
    \bottomrule
    \end{tabular}
    \caption{The lowest excitation energy ($\omega$) of the PNA-in-water system, calculated with MLCC2-in-HF, MLCC2-in-MLHF, MLCCSD-in-HF, and MLCCSD-in-MLHF using the frozen core approximation. The atoms within a radius of $\SI{6}{\angstrom}$ are included in the MLCC calculation and the atoms within a radius of $\SI{3.5}{\angstrom}$ are treated with the higher level coupled cluster method (CC2 or CCSD). In the MLHF reference calculation, the atoms within a radius of $\SI{10}{\angstrom}$ are active. The aug-cc-pVDZ basis is used on the atoms that are included in the MLCC calculation, and cc-pVDZ is used on the remaining atoms. The wall times for the reference calculation ($t^\mrm{Ref}$) and the MLCC calculation  ($t^\mrm{MLCC}$) are also given. The calculations were performed on two Intel Xeon Gold 6152 processors with 44 threads and 1.4 TB memory available. The peak memory usage (PMU) is given in GB.}
    \label{tab:big_pna_2}
\end{table}

We have also performed MLCC calculations on the PNA-in-water system in Figure \ref{fig:pna_499} with larger basis sets. 
In Table \ref{tab:big_pna_2}, we present timings for MLCC2-in-HF/MLHF and MLCCSD-in-HF/MLHF calculations with $r_{\text{CCS}}=\SI{6.0}{\angstrom}$ and $r_{\text{CC2/CCSD}}=\SI{3.5}{\angstrom}$.
The aug-cc-pVDZ basis is used for all atoms included in CC active region, and cc-pVDZ is used on the remaining atoms. In total, there are $12669$ AOs and $1498$ MOs \revS{in the coupled cluster calculation}. The Cholesky decomposition is performed with MO-screening using a threshold of $10^{-4}$. \revS{For the reference calculations, a gradient threshold of $10^{-6}$ is used.}

Comparing Tables \ref{tab:CD_pna_water} and \ref{tab:big_pna_2}, we see that the MLCCSD-in-HF excitation energies do not change significantly with a larger basis and an increased $r_{\text{CCS}}$.
For the calculations presented in Table \ref{tab:big_pna_2}, the reference calculation \revE{is the most expensive step}. 
Since the active region of the MLHF calculation is large ($\SI{10.0}{\angstrom}$), we do not obtain large savings using an MLHF reference. However, this can be achieved by reducing $r_{\text{HF}}$. Furthermore, MLHF is applicable for systems where standard Hartree-Fock is not computationally feasible.
\revS{The CC2-in-HF calculation for this system, with a CC2 radius of $\SI{6}{\angstrom}$, yields $\omega=\SI{3.732}{\eV}$. Hence, the error of using MLCC2, compared to CC2, is approximately $\SI{0.1}{\eV}$.
The effect of extending the CCS radius to $r_{\text{CCS}}=\SI{8.0}{\angstrom}$ is to increase the excitation energy by $\SI{0.003}{\eV}$ to $\omega = \SI{3.824}{\eV}$.}

Solvation effects can be estimated by performing calculations on a series of snapshots from a molecular mechanics simulation, for instance using the QM/MM approach for the individual snapshots, such as in  Ref.~\citenum{giovannini2019fqfmulinear}.
The calculations in this paper demonstrate that a fully quantum mechanical approach---MLCC-in-HF and MLCC-in-MLHF---can be used to determine such solvation effects.
\revS{For the former of these approaches, the Hartree-Fock calculation is likely to be the time limiting step.}

\begin{figure*}
    \centering
    \includegraphics[width=0.6\textwidth]{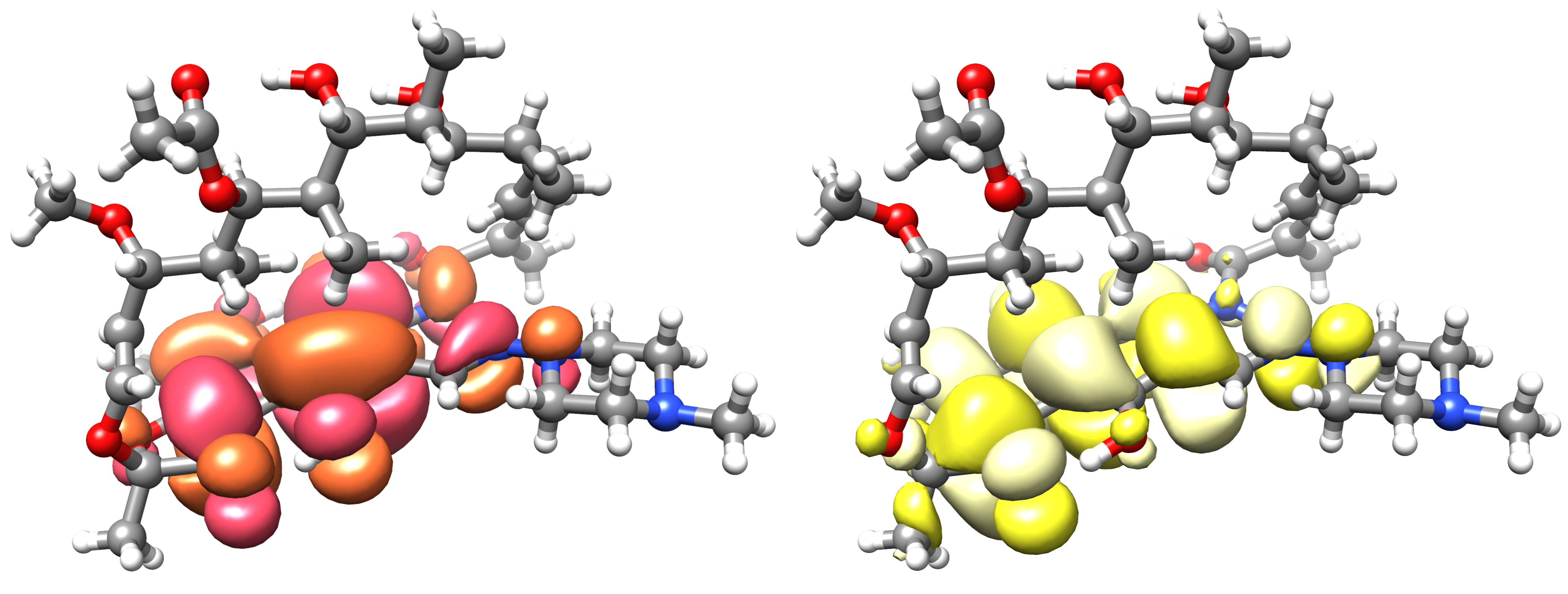}
    \includegraphics[width=0.3\textwidth]{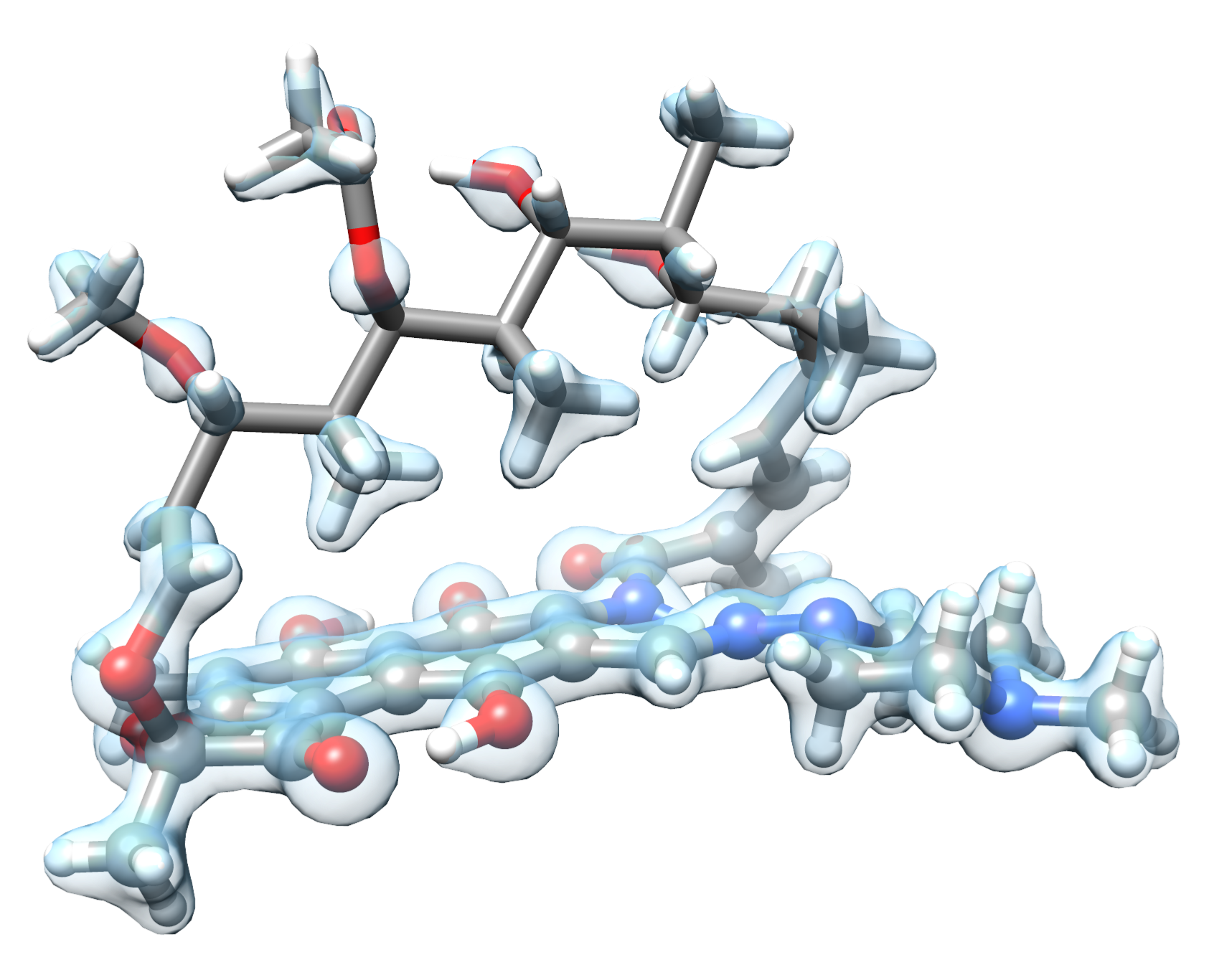}
    \caption{\revS{Dominant occupied (left) and virtual (center) NTOs (CCS/aug-cc-pVDZ) of the lowest excited state of rifampicin. The Hartree-Fock density of the active occupied MOs used in the CC-in-HF/aug-cc-pVDZ and MLCC-in-HF/aug-cc-pVDZ calculations (right); the active atoms are illustrated with ``ball-and-stick'' and the inactive atoms as ``sticks''.}}
    \label{fig:rifampicin_orbs_and_dens}
\end{figure*}

\begin{table}[]
    \centering
    \begin{threeparttable}
    \begin{tabular}{l c c c }
    \toprule
    Method 
    & $n_o^{\text{a}}$ 
    & $n_v^{\text{a}}$ 
    & $\omega~[\si{\eV}]$ \\
    \midrule
    {MLCC2}
    & 60 & 600 & 2.65  \\
    {MLCC2-in-HF}
    & 60 & 600 &  2.77 \\
    \midrule
    CC2 & 161 & 1645 & 2.57  \\
    CC2-in-HF & 131 & 981 & 2.70  \\
    \midrule
    {MLCCSD}
    & 50 & 500 & 3.02 \\
    {MLCCSD-in-HF}
    & 50 & 500 & 3.13\\
    MLCCSD-in-HF\tnote{$\dagger$} & 50 & 500 & 3.16 \\
    \bottomrule
    \end{tabular}
    \begin{tablenotes}\item [$\dagger$] aug-cc-pVTZ on active atoms \end{tablenotes}
    \end{threeparttable}
    \caption{\revS{X and X-in-HF calculations for the lowest excitation energy in rifampicin, with X = \{MLCC2, CC2, MLCCSD\}. The aug-cc-pVDZ basis is used, unless otherwise is stated. The active atoms in the CC and MLCC calculations were selected by hand (see Figure \ref{fig:rifampicin_orbs_and_dens}). CNTOs were used to partition the orbitals in the MLCC calculations.}}
    \label{tab:rifampicin_all}
\end{table}
\revS{
The MLCC-in-HF and MLCC-in-MLHF approaches are not only applicable to solute-solvent systems\revE{. They} can also be used for large molecules.
As a proof of concept, we present MLCC-in-HF calculations for the lowest excitation in rifampicin in Table \ref{tab:rifampicin_all}. In Figure \ref{fig:rifampicin_orbs_and_dens}, we have plotted the NTOs from a CCS/aug-cc-pVDZ calculation. The excitation is seen to be located in a subregion of the molecule. \revE{It} can therefore be treated with CC-in-HF or MLCC-in-HF. In Figure \ref{fig:rifampicin_orbs_and_dens}, we have also plotted the Hartree-Fock density of the active occupied orbitals treated with CC or MLCC. 
The active atoms in the CC and MLCC calculations were selected by hand by inspecting the NTOs. CNTOs were used to partition the orbitals in the case of MLCC2/MLCCSD.
The shift observed by going from  $X$ to $X$-in-HF is about $\SI{0.1}{\eV}$ in all the presented calculations. It should be noted that this system is too small to be suitable for CC-in-HF and MLCC-in-HF, but is chosen because the reference CC2 calculations are available.  The MLCC2 or MLCCSD methods are preferable for systems of this size since, as can be seen from Tables \ref{tab:rifampicin} and \ref{tab:rifampicin_mlccsd}, these calculations can be performed with ease.
}

\section{Concluding remarks}
In this paper, we have demonstrated the computational savings that can be obtained with MLCC2 and CCS/CCSD MLCCSD. These multilevel methods can be used for systems that are too large to be described at the CC2 and CCSD level. However, the MLCC2 and MLCCSD models are limited by the underlying scaling of the lower-level coupled cluster method (CCS). We have therefore presented a framework of reduced-space MLCC that can be used for systems with several thousand AOs.
In this \revL{layered approach},
MLCC is only applied to a restricted region of the molecular system; the environment is optimized with Hartree-Fock, or multilevel Hartree-Fock, and only contributes to the MLCC calculation through the Fock matrix. 
Efficient implementation of this framework requires careful handling of the electron repulsion integrals. 
We have implemented a direct construction of MO Cholesky vectors that reduces the storage requirement to $\mathcal{O}(N_\mathrm{AO} n_\mathrm{MO}^2)$. 
With an additional screening during the Cholesky decomposition algorithm, we further reduce this requirement to $\mathcal{O}(n_\mathrm{MO}^3)$\revE{, making the storage requirement independent of the size of the environment.}
Exploiting the Cholesky factorization in this manner, we can handle systems with several thousand basis functions using existing MLCC implementations.
The MLCC-in-HF/MLHF framework is 
therefore
suited to 
accurately model
solvation effects on intensive properties on the solute. It can also be used for chromophores in biomolecules. 

\section{Acknowledgements}
We thank Rolf H. Myhre for insightful discussions and for his work on optimization in the e$^T$ program, where the MLCC code is implemented.
We also thank Ida-Marie H{\o}yvik and Tommaso Giovannini for helpful discussions.
We acknowledge computing resources through UNINETT Sigma2 - 
the National Infrastructure for High Performance Computing and Data Storage in Norway,
through project number NN2962k. 
We acknowledge funding from the Marie Sk{\l}odowska-Curie European Training Network ``COSINE - COmputational Spectroscopy In Natural sciences and Engineering'', Grant Agreement No. 765739 and the Research Council of Norway through FRINATEK projects 263110 and 275506.

\section{Appendix A}
We use the following notation for the MLCC2 and MLCCSD equations:
indices $a, b, c, ...$ denote active virtual orbitals, $A, B, C, ...$ unrestricted virtual orbitals, $i, j, k, ...$ active occupied orbitals, $I, J, ...$ unrestricted occupied orbitals, $p, q, r, ...$ general active orbitals, and $P, Q, R, ...$ general unrestricted orbitals. The index $K$ is used to denote Cholesky vectors.
We also define $n_o$ and $n_v$ as the number of active occupied and active virtual orbitals, respectively, and $N_o$ and $N_v$  as the total number of occupied and virtual orbitals. We also use $N_{\MO}$ and $N_J$ for the number of MOs and Cholesky vectors, respectively.
We have adopted the Einstein notation with implicit summation over repeated indices.
In the following screening considerations, we assume a fixed active space and and expanding inactive space.

The electron repulsion integrals are Cholesky decomposed,
\begin{align}
    g_{PQRS} = L^K_{PQ}L^K_{RS}.
\end{align}
The Cholesky vectors are stored in both the MO and the $X_1$-transformed basis. In general, $X_1$-transformed quantities are denoted with tilde, e.g.:
\begin{align}
     \tilde{g}_{PQRS} = \tilde{L}^K_{PQ}\tilde{L}^K_{RS}.
\end{align}
When no indices are restricted to the active space, the construction of $\b{g}$ or $\b\Tg$ from $\b{L}$ or $\b\TL$, respectively, is an $\mathcal{O}(N^5)$ operation.

\subsection{The ground state equations and the correlation energy}
Solving the projected coupled cluster equations,
\begin{align}
    \Omega_{\mu} = \Tbraket{\mu}{\Hsim}{\HF} = 0,\label{eq:omega_general}
\end{align}
entails the iterative construction of the $\b{\Omega}$-vector, the iterative construction of the Fock matrix in the $X_1$-transformed basis ($\b{\TF}$), and the calculation of the correlation energy. The correlation energy is computed in every iteration, however, this is not necessary as convergence can be determined purely from the norm of $\b{\Omega}$.
 
The Fock matrix in the $X_1$-transformed basis is given by
\begin{align}
    \TF_{PQ} = 2\Tg_{PQII} - \Tg_{PIIQ} = 2 \TL^K_{PQ}\TL^K_{II} - \TL^K_{PI}\TL^K_{IQ},
\end{align}
and its construction, in terms of the Cholesky vectors, scales as $N_{\mrm{MO}}^2 N_O N_J$ ($\mathcal{O}(N^4)$). However, depending on the coupled cluster model, only certain subblocks of $\b\TF$ are needed to solve \eqref{eq:omega_general}.

In MLCCSD and MLCC2, the correlation energy is given by
\begin{align}
    E_{\mrm{correlation}} = \x{A}{I}\x{B}{J}L_{IAJB}  + \x{ab}{ij}L_{iajb},\label{eq:mlcc_energy}
\end{align}
where we have introduced $L_{PQRS} = 2g_{PQRS} - g_{PSRQ}$. The last term in eq \eqref{eq:mlcc_energy} is restricted to the active space. The first term is calculated according to
\begin{align}
    \x{A}{I}\x{B}{J}L_{IAJB} = 2(L^K_{IA}\x{A}{I})\cdot(L^K_{JB}\x{B}{J}) - (L^K_{IB}\x{B}{J})\cdot(L^K_{JA}\x{A}{I}),
\end{align}
which scales as $N_VN_O^2N_J$ ($\mathcal{O}(N^4)$), avoiding the $\mathcal{O}(N^5)$ integral constructions.
%%%%%%%%%%%%%%%%%%%%%%%%%%%%%%%%%%%%%%%%%
%%%%%%%%%%%%%%%%%%%%%%%%%%%%%%%%%%%%%%%%%
\subsubsection{The MLCC2 $\b{\Omega}$-vector}
%%%%%%%%%%%%%%%%%%%%%%%%%%%%%%%%%%%%%%%%%
%%%%%%%%%%%%%%%%%%%%%%%%%%%%%%%%%%%%%%%%%
In MLCC2, the cluster operator is given by
\begin{align}
    X = X_1 + S_2
\end{align}
$\b{\Omega}$-vector becomes
\begin{align}
        \Omega_{\mu_1} = &\Tbraket{\mu_1}{\HTone + [\HTone, S_2]}{\HF} = 0\label{eq:mlcc2_omega1}\\
        \Omega_{\mu_2} = &\Tbraket{\mu_2^{S}}{\HTone + [F, S_2]}{\HF} = 0.\label{eq:mlcc2_omega2}
\end{align}
Eq \eqref{eq:mlcc2_omega2} can be solved analytically for the $s$-amplitudes:
\begin{align}
  \s{ab}{ij} = -\frac{\Tg_{aibj}}{F_{aa} + F_{bb} - F_{ii} - F_{jj}},  
\end{align}
where $F_{PQ}$ are elements of the Fock matrix.
The $\b{\Omega}$-vector is coded as
\begin{align}
    \Omega_{AI} = {\TF}_{AI} +
    \Big((\u{bc}{ij}\TL^K_{jc})\TL^K_{Ab}\Big)\delta_{Ii}
    - \Big(\u{ab}{jk}\Tg_{kbjI}\Big)\delta_{Aa}
    + \Big(\u{ab}{ij}\TF_{jb}\Big)\delta_{AI,ai}
    \label{eq:mlcc2_omega_codable}
\end{align}
where $\u{ab}{ij} = 2\s{ab}{ij} - \s{ab}{ji}$. The calculation of eq \eqref{eq:mlcc2_omega_codable} entails two contractions scaling as $\mathcal{O}(N^2)$ and two contractions scaling as $\mathcal{O}(N)$.

%%%%%%%%%%%%%%%%%%%%%%%%%%%%%%%%%%%%%%%%%
%%%%%%%%%%%%%%%%%%%%%%%%%%%%%%%%%%%%%%%%%
\subsubsection{The MLCCSD $\b{\Omega}$-vector}
%%%%%%%%%%%%%%%%%%%%%%%%%%%%%%%%%%%%%%%%%
%%%%%%%%%%%%%%%%%%%%%%%%%%%%%%%%%%%%%%%%%
In MLCCSD, the cluster operator is given by
\begin{align}
    X = X_1 + T_2
\end{align}
$\b{\Omega}$-vector is given by
\begin{align}
        \Omega_{\mu_1} = &\Tbraket{\mu_1}{\HTone + [\HTone, T_2]}{\HF} = 0\label{eq:mlcc2_omega1}\\
        \Omega_{\mu_2} = &\Tbraket{\mu_2^{T}}{\HTone + [\HTone, T_2] + \tfrac{1}{2}[[\HTone, T_2],T_2]}{\HF} = 0.\label{eq:mlcc2_omega2}
\end{align}
${\Omega}_{\mu_1}$ is the same as in MLCC2, but with $t$-amplitudes in place of $s$-amplitudes.
${\Omega}_{\mu_2}$ is given by:
\begin{align}
    \begin{split}
    \Omega_{aibj} =&\;\Tg_{aibj} + \Tg_{acbd}\t{cd}{ij} + \t{ab}{kl}(\Tg_{kilj} + \t{cd}{ij}\Tg_{kcld})\\
    &- \tfrac{1}{2}\t{bc}{kj}Y_{aick} - \t{bc}{ki}Y_{ajck}\\
    &- \tfrac{1}{2}\u{bc}{jk}\Tg_{acki} + \tfrac{1}{4}\u{bc}{jk}\TL_{ldkc}\u{ad}{il} + \u{bc}{jk}g_{aikc}\\
    &+ \t{ac}{ij}(\TF_{bc} - \Tg_{ldkc}\u{bd}{kl}) -  \t{ab}{ik}(\TF_{kj} - \Tg_{ldkc}\u{dc}{lj})
    \end{split}
\end{align}
where $Y_{aick} = g_{kiac} - \tfrac{1}{2}\t{cd}{ij}\Tg_{kcld}$. All orbital indices are restricted to the active space and only the integral construction scales with the system (linear scaling, $\mathcal{O}(N)$).

\subsection{Jacobian transformation}
The linear transformation by the Jacobian matrix,
\begin{align}
   \b \sigma = \b A \b c,
\end{align}
must be calculated in order to obtain excitation energies in coupled cluster theory.

\subsubsection{MLCC2 Jacobian transformation}
The MLCC2 Jacobian matrix is given by
\begin{equation}
  \b A^{\mrm{MLCC2}} = 
  \begin{pmatrix}
    \Tbraket{\mu_1}{[\HTone,\tau_{\nu_1}]+ [[\HTone, \tau_{\nu_1}], S_2]}{\HF} &
    \Tbraket{\mu_1}{[\HTone, \tau_{\nu_2^{S}}]}{\HF}  \\
    \Tbraket{\mu_2^{S}}{[\HTone, \tau_{\nu_1}]}{\HF} &
    \Tbraket{\mu_2^{S}}{[F, \tau_{\nu_2^{S}}]}{\HF} 
  \end{pmatrix}.
\end{equation}
The block $\Tbraket{\mu_2^{S}}{[F, \tau_{\nu_2^{S}}]}{\HF}$ reduces to $\epsilon_{\mu_2^{S}}\delta_{\mu_2^{S}\nu_2^{S}}$ in the semicanonical basis, where $\epsilon_{aibj} = F_{aa} + F_{bb} - F_{ii} - F_{jj}$. 

The terms of the singles part of the transformed vector are:
\begin{align}
\begin{split}
    \sigma^{\mrm{MLCC2}}_{AI} =& 
    \TF_{AB}c_{BI} - \TF_{JI}c_{AJ} + % CCS A1
    2(\TL^K_{JB}c_{BJ})\TL^K_{AI} - (\TL^K_{AB}c_{BJ})\TL^K_{JI}\\ % CCS B1
    &+ 2\Big((\TL^K_{JB}c_{BJ})\TL^K_{kc}\Big)\u{ac}{ik}\delta_{AI,ai} 
    - \Big((\TL^K_{kB}c_{BJ})\TL^K_{Jc}\Big)\u{ac}{ik}\delta_{AI,ai}\\ % MLCC2 A1
    & - X_{Ji}c_{AJ}\delta_{Ii} - Y_{aB}c_{BI}\delta_{Aa} \\ % MLCC2 A1
    & + \TF_{jb}(2c_{aibj} - c_{ajbi}) - L_{jbki}c_{akbj} % MLCC2 B1
\end{split}\label{eq:mlcc_A_}
\end{align}
The intermediates $\b{X}$ and $\b{Y}$,
\begin{align}
    X_{Ji} &= \Tg_{Jbkc} \u{cb}{ki}\\
    Y_{aB} &= \Tg_{kcjB}\u{ac}{jk} ,
\end{align}
are calculated once before the iterative loop.
The fourth term of eq \eqref{eq:mlcc_A_} scales as $\mathcal{O}(N^4)$, and it is the steepest scaling term. 
Additionally, there are several contractions that scale as $\mathcal{O}(N^3)$.
If we compare to the transformation by the CCS Jacobian,
\begin{align}
    \sigma^{\mrm{CCS}}_{AI} =& 
    \TF_{AB}c_{BI} - \TF_{JI}c_{AJ} + % CCS A1
    2(\TL^K_{JB}c_{BJ})\TL^K_{AI} - (\TL^K_{AB}c_{BJ})\TL^K_{JI}, % CCS B1
\end{align}
we see that the steepest scaling term enters at the CCS level of theory.

The terms of the doubles part of the transformed vector are:
\begin{align}
    \sigma^{\mrm{MLCC2}}_{aibj} = (\TL^K_{bC}c_{Cj})\TL^K_{ai} - \Tg_{Kjai}c_{bK} + \epsilon^{ab}_{ij}c_{aibj}
\end{align}
Its construction entails two $\mathcal{O}(N^2)$ (term 1 and the construction of the integrals used in term 2) and two $\mathcal{O}(N)$. 

\subsubsection{MLCCSD CCS/CCSD Jacobian transformation}
The CCS/CCSD MLCCSD Jacobian matrix is given by
\begin{align}
  &\b A^{\mrm{MLCCSD}} =
  &\begin{pmatrix}
    \Tbraket{\mu_1}{[{\HTone},\tau_{\nu_1}]+ [[{\HTone}, \tau_{\nu_1}], T_2]}{\HF} &
    \Tbraket{\mu_1}{[{\HTone}, \tau_{\nu_2^{\mrm{T}}}]}{\HF} \\
    \Tbraket{\mu_2^{{T}}}{[{\HTone},\tau_{\nu_1}]+ [[{\HTone}, \tau_{\nu_1}], T_2]}{\HF} &
    \Tbraket{\mu_2^{{T}}}{[{\HTone},\tau_{\nu_2^{{T}}}]+ [[{\HTone}, \tau_{\nu_2^{{T}}}], T_2]}{\HF}
  \end{pmatrix}.
\end{align}
The singles part of $\b{\sigma}^{\mrm{MLCCSD}}$ is the same as in MLCC2, see eq \eqref{eq:mlcc_A_}. The doubles part is given by:
\begin{align}
\begin{split}
    \sigma_{aibj}^{\text{MLCCSD}} 
  & = \frac{\mathcal{P}^{ab}_{ij}}{1+\delta_{aibj}}\bigg(
    (\TL^K_{bC}c_{Cj})\TL^K_{ai} - \Tg_{Kjai}c_{bK} \\ % A2 (CC2) 
  & - ({\TF}_{Kc} t_{ij}^{ac}) c_{bK} % B2 
    - ({\TF}_{kC} t_{ik}^{ab}) c_{Cj} \\ 
  & + [\b{Y1}]_{Kjai} c_{bK} %C2 
    + [\b{Y2}]_{Kjbi} c_{aK}\\
  & + ({\Tg}_{ljkC} c_{Ci}) t_{lk}^{ba} %7
    - ({\TL}_{ljKC}c_{CK}) t_{il}^{ab} \\%8
  & - [\b{Y3}]_{Kibj} c_{aK} %D2 %9
    - ({\Tg}_{kDbc}c_{Di}) t_{kj}^{ac} %10
    - ({\Tg}_{kcbD}c_{Dj}) t_{ik}^{ca} \\%11
  & + ({\TL}_{kcbD}c_{Dj}) t_{ik}^{ac} %12
    + ({\TL}_{KCbd}c_{CK}) t_{ij}^{ad}\\ %13
  & + [\b{Y4}]_{ckbj} (2\tilde{c}_{aick} - \tilde{c}_{akci}) % E2 %14
    - (\sum_{ckl}{\TL}_{kcld} \tilde{c}_{blck}) t_{ij}^{ad} %F2 %15
    - ({\TL}_{kcld}\tilde{c}_{ckdj}) t_{il}^{ab} \\ %16
  & - [\b{Y5}]_{ckbj} \tilde{c}_{aick} %G2  %17
    - [\b{W1}]_{cb} \tilde{c}_{aicj} %18
    - [\b{W2}]_{kj} \tilde{c}_{aibk}\\  %19
  & + [\b{Y6}]_{aick} \tilde{c}_{bkcj} %H2 %20
    + [\b{Y7}]_{ajck}\tilde{c}_{bkci}\\ %21
  & + {\TF}_{bc} \tilde{c}_{aicj} - {\TF}_{kj} \tilde{c}_{aibk} \\%I2 %22-23
  & + {\Tg}_{bjkc}(2\tilde{c}_{aick} - \tilde{c}_{akci}) 
    - {\Tg}_{bckj}\tilde{c}_{aick} 
    - {\Tg}_{kibc}\tilde{c}_{akcj}\\
  & + [\b{Y8}]_{klij} \tilde{c}_{akbl} % J2 %27
    + ({g}_{kcld} \tilde{c}_{cidj}) t_{kl}^{ab} \\ %28
  & + {\Tg}_{kilj} \tilde{c}_{akbl} % K2 %29
    + {\Tg}_{acbd} \tilde{c}_{cidj}\bigg) %30
\end{split}\label{eq:sigma_mlccsd}
\end{align}

The contractions in eq \eqref{eq:sigma_mlccsd} scale as either $\mathcal{O}(0)$, $\mathcal{O}(N)$, or $\mathcal{O}(N^2)$. 
Additionally, the integral constructions scale as $\mathcal{O}(N)$, $\mathcal{O}(N^2)$, or $\mathcal{O}(N^3)$.

The intermediates are calculated once before the iterative loop, and are defined as:
\begin{align}
    [\b{Y1}]_{Kjai} & = {\Tg}_{Kjlc} t_{li}^{ac} \\
    [\b{Y2}]_{Kjbi} & = {\Tg}_{ljKc} t_{li}^{bc} \\
    [\b{Y3}]_{Kibj} & = {\Tg}_{Kcbd} t_{ij}^{cd} \\
    [\b{Y4}]_{ckbj} & = {\TL}_{kcld} t_{jl}^{bd} \\
    [\b{Y5}]_{ckbj} & = {\TL}_{kcld} t_{lj}^{bd} \\
    [\b{Y6}]_{aick} & = {\Tg}_{ldkc} t_{il}^{da} \\
    [\b{Y7}]_{ajck} & = {\Tg}_{kdlc} t_{jl}^{da} \\
    [\b{Y8}]_{klij} & = {\Tg}_{kcld} t_{ij}^{cd} \\
    [\b{W1}]_{cb}   & = {\TL}_{kdlc} t_{lk}^{bd} \\
    [\b{W2}]_{kj}   & = {\TL}_{lckd} t_{lj}^{cd}
\end{align}

\bibliography{main}% Produces the bibliography via BibTeX.

\begin{figure}
    \centering
    \includegraphics{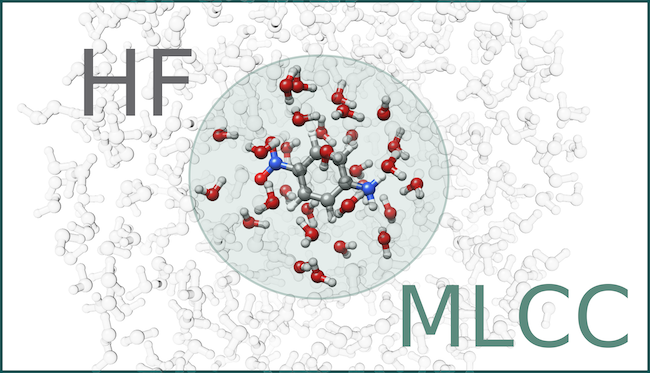}
    \caption{For Table of Contents Only}
\end{figure}

\end{document}